\documentclass[twoside,8pt]{article}
\usepackage{epsfig}

\newcommand{\be}{\begin{equation}}
\newcommand{\ee}{\end{equation}}
\newcommand{\bea}{\begin{eqnarray}}
\newcommand{\eea}{\end{eqnarray}}

\begin{document}

%\title{Equation of state effects on the lower limit of the maximum mass of neutron stars }

\title{ Effects of the equation of state on the core-crust interface of  slowly rotating neutron stars}

\author{
L. Tsaloukidis$^1$, Ch. Margaritis$^1$ and Ch.C. Moustakidis$^{1,2}$\\
$^1$Department of Theoretical Physics, Aristotle University of
Thessaloniki,\\  54124 Thessaloniki, Greece\\
$^{2}$Theoretical Astrophysics, Eberhard Karls University of T\"{u}bingen,\\
 T\"{u}bingen 72076, Germany }

\maketitle

\begin{abstract}

We systematically  study the symmetry energy  effects of the transition  density $n_{\rm t}$ and the transition pressure $P_{\rm t}$ around the crust-core interface of a neutron star in the framework of the dynamical and the thermodynamical method respectively. We employ both the parabolic approximation and  the full expansion, for the definition of the symmetry energy. We use  various theoretical  nuclear models, which are suitable for reproducing
the bulk properties of nuclear matter at low densities, close to saturation density as well as the maximum observational neutron star mass. Firstly we derive and present an approximation for the transition pressure $P_{\rm t}$ and crustal mass $M_{\rm crust}$. Moreover, we derive  a model-independent correlation between $P_{\rm t}$ and the slope parameter $L$ for a fixed value of the symmetry energy at the saturation density. Secondly, we explore the effects of the Equation of State (EoS) on  a few astrophysical applications which are sensitive to the values of $n_{\rm t}$ and $P_{\rm t}$ including  neutron star oscillation frequencies, thermal relaxation of the crust,  crustal fraction of the moment of inertia and the r-mode instability window  of a rotating neutron star.  In particular, we employ the Tolman VII solution of the TOV equations to derive analytical expressions for the critical frequencies and the relative time scales, for the r-mode instability, in comparison with the numerical predictions.  In the majority of the applications, we found that the above quantities  are sensitive mainly  to the applied  approximation  for  the symmetry energy (confirming previous results). There is also a dependence on the used method (dynamical or thermodynamical). The above findings lead us to claim that the determination of  $n_{\rm t}$ and $P_{\rm t}$ must be reliable and accurate before  they are used to constrain relevant neutron star properties.

\vspace{0.3cm}

PACS number(s):
26.60.-c, 26.60.Kp, 21.65.Ef, 26.60.Cj \\

Keywords: Neutron stars; Nuclear equation of state; Nuclear
symmetry energy; Crust-core interface; Dynamical method; Neutron star instabilities
\end{abstract}

%%%%%%%%%%%%%%%%%%%%%%%%%%%%%%%%%%%
%\section{Introduction}
%%%%%%%%%%%%%%%%%%%%%%%%%%%%%%%%%%%%
%\end{document}

%%%%%%%%%%%%%%%%%%%%%%%%%%
\section{Introduction}
%%%%%%%%%%%%%%%%%%%%%%%%%%%
Neutron stars (NSs) are the most compact stellar objects in the universe,
which makes them extraordinary astronomical laboratories for the physics
of dense nuclear matter~\cite{Shapiro-83,Glendenning-2000,Haensel-07}. Very recently, the detection of
gravitational waves from the merger of two neutron stars, in a binary
neutron-star system, opened a new powerful window to the exploration of
the physics of NSs~\cite{Abbott-017,Ligo-017}. Particularly, many of the  static properties  as well as dynamical processes  of  neutron stars  are sensitively dependent  on the employed equation of state~\cite{Petchik-95a,Lattimer-012,Watts-016,Oertel-017,Baldo-16,Giuliani-14,Graber-2017,Bao-2008,Epja-014}. However, the knowledge of the  equation of state, especially at high densities, is very uncertain and consequently the relevant predictions and estimations suffer from uncertainties. On the other hand, for low densities (close to the saturation density of symmetric nuclear matter) the EoS is well constrained and the relevant predictions are more reliable. This prediction includes the crust-core interface which is the main subject of the present study.

The interior of a neutron star is divided into the outer core and the inner one. It  has a radius of approximately 10-14 km and contains most of the star's mass~\cite{Petchik-95a}. The crust, with a thickness of about $10\%$ of the total radius contains only a few percent of the total mass. It  can also be divided into an outer and an inner part. The equation of state of neutron-rich nuclear matter is the important ingredient among all the bulk properties of NS  in the study of both  the core and the crust. In particular, the implementation of the EoS  predicts the  location of the inner edge of a neutron star crust.
The inner crust comprises the outer region from the density at which neutrons drip out of nuclei to the inner edge, separating the solid crust
from the homogeneous liquid core. At the inner edge, in fact, a phase transition occurs from the high density homogeneous matter to the inhomogeneous one at lower densities. It was found that the transition density is related to some finite nuclei properties including neutron-skin, dipole polarizability e.t.c~\cite{Paar-014,Centelles-09,Horowitz-2001}.

The baryon transition density $n_{\rm t}$ at the inner edge is uncertain due to our insufficient knowledge of the EoS of neutron-rich nuclear matter. In addition, the determination of the transition density $n_{\rm t}$ itself is a very complicated problem because the inner crust
may have  an intricate  structure. A well established approach is to find the density at which the uniform liquid first becomes
unstable against small-amplitude density fluctuations, indicating the formation of nuclear clusters. This approach
includes the dynamical method~\cite{Baym-71,Pethick-95b,Oyamatsu-07,Ducoin-07,Xu-09,Lattimer-13,Feng-17}, the thermodynamical one~\cite{Kubis-07a,Kubis-07b,Moustakidis-010,Moustakidis-012}  and the random phase approximation (RPA)~\cite{Horowitz-2001,Carriere-03}. Recently, a method to determine the transition density in the framework of the unified equation of state, has been presented in Ref.~\cite{Fortin-016}.

The structure of the crust  as well as some dynamical processes are affected appreciably by
the location of the crust-core interface.
Firstly, if the transition density $n_{\rm t}$ is sufficiently high, it is possible for nonspherical phases, with
rod- or plate-like nuclei, to occur before the nuclei dissolve~\cite{Oyamatsu-93,Caplan-017}. If $n_{\rm t}$ is relatively low, then the matter undergoes a direct transition from spherical nuclei to uniform nucleonic fluid. In general, the values of the  transition  density are related to the existence of the nuclear pasta, including various phases i.e.  droplet, rod, slab, tube and bubble (see Refs.~\cite{Fortin-016,Sharma-015} for a recent study) but we will not be  studying  this issue in the present work.  The pulsar glitches (sudden discontinuities in the spin-down of a pulsar)  are related to the crustal fraction of the moment of inertia~\cite{Haskell-015,Ho-015,Delsate-016}. Moreover, the frequencies of a class of  neutron star oscillations, which can be detected from observations of quasi-periodic oscillations in the X-ray emissions, are dependent  on the transition density between crust and core~\cite{Lattimer-07,Samuelsson-07,Sotani-017}. In the dynamical process of the neutron star cooling, the thermal relaxation of the crust is sensitive to the crust radius~\cite{Lattimer-07,Gnedin-01,Lattimer-94}. In addition, concerning  the r-mode instability condition, the critical angular velocity depends appreciably on the core radius,  the transition density and the energy density~\cite{Lidblom-2000,Andersson-1998,Friedman-98,Friedman-99,Andersson-2001,Andersson-2003,
Kokkotas-99,Andersson-99,Moustakidis-015}.

The motivation of the present work is twofold. Firstly, in the framework of the dynamical and  the  thermodynamical method we calculate the transition density and the corresponding  pressure using various nuclear models. In particular we explore the effects of the contribution of the Coulomb and the density gradient terms  on the determination of $n_{\rm t}$ and $P_{\rm t}$ and consequently on some neutron star properties, while examining   in parallel how  the nuclear symmetry energy affects the above mentioned values. Secondly, we concentrate our study  mainly on the  error which can be introduced by employing the well known parabolic approximation for the symmetry energy, not only on the values of $n_{\rm t}$ and $P_{\rm t}$ but also on the predictions of some neutron star observable properties. We exhibit the necessity to implement both the dynamical method  and the full approximation for the symmetry energy in order to get reliable predictions.

Moreover, we provide analytical expressions for the mass of the crust  $M_{\rm crust}$ and also for the transition pressure $P_{\rm t}$. A semi-analytical expression, based on theoretical and empirical arguments,  has been derived  and presented for the $P_{\rm t}$.  In particular,   considering fixed values of the symmetry energy at the saturation density, we arrive at a model-independent relation between $P_{\rm t}$ and the slope parameter $L$. Finally, we employ the Tolman VII analytical  solution of the TOV equations and we derive analytical expressions for the time scales and frequencies related with the r-mode instabilities (which are sensitive to the crust-core interface). The proposed approximation has been proved to be  very accurate,  providing some useful analytical relations suitable for astrophysical applications.

The article is organized as follows. In Sec.~2, we present the basic formalism of the dynamical method and also all the key expressions needed to calculate the transition density and the corresponding pressure including the nuclear symmetry energy formalism. The nuclear models used in the present study are presented in Sec.~3.
In Sec.~4 we present  applications of the methods in various astrophysical issues.  Our results are presented and discussed in Sec.~5 while Sec.~6 summarizes the present work.

%%%%%%%%%%%%%%%%%%%%%%%%%%%%%%%%%%%%%%%%%%%%%%%%%
\section{The dynamical method formalism}
%%%%%%%%%%%%%%%%%%%%%%%%%%%%%%%%%%%%%%%%%%%%%%%
The study of the instability of $\beta$ stable nuclear matter is based on the variation of the total energy density, in the framework of the Thomas-Fermi approximations (see the innovative work by Baym, Bethe and Pethick~\cite{Baym-71}). In the  dynamical method,  compared to the thermodynamical one,  effects from inhomogeneities of the
density and  the Coulomb interaction have also been included. The starting point of this method  is the consideration of  small
sinusoidal variations in the neutron, proton and electron densities defined as  $\delta n_n({\bf r})$, $\delta n_p({\bf r})$ and $\delta n_e({\bf r})$.
%\begin{equation}
%\delta_i({\bf k},{\bf r})=\delta_i({\bf k})e^{i{\bf k}\cdot {\bf r}}, \quad i=n,p,e
%\label{dn-1}
%\end{equation}
%
%where $\delta_i({\bf k})$ is the density in momentum space.
The onset  of instability occurs when  the total energy in the presence of density inhomogeneity is lower than the energy of the uniform liquid.
In particular, the expansion of the total energy
up to second order in the variation of the densities leads to~\cite{Baym-71,Pethick-95b}
\begin{eqnarray}
{\cal E}-{\cal E}_0&=&\frac{1}{2}\sum_{i,j}\int \frac{\delta^2{\cal E}}{\delta n_i({\bf k})\delta n^*_j({\bf k})}\delta n_i({\bf k})\delta n^*_j({\bf k})\frac{d{\bf k}}{(2\pi)^3}\nonumber\\
&=&\frac{1}{2}\int { U}_{\rm dyn}(k,n)|\delta n_p({\bf k})|^2  \frac{d{\bf k}}{(2\pi)^3},
\end{eqnarray}
where ${\cal E}_0$ is the energy of the uniform phase and $\delta n_{i}({\bf k})$ is the density in momentum space. The onset of instability will occur if the total energy ${\cal E}$, in the presence of the density inhomogeneity, is lower than ${\cal E}_0$.  ${ U}_{\rm dyn}(k,n)$ is the so-called effective interaction between protons given by~\cite{Baym-71,Pethick-95b}
\begin{eqnarray}
{ U}_{\rm dyn}(k,n)=\left(\frac{\partial \mu_p}{\partial n_p}+2D_{pp}k^2+\frac{4\pi e^2}{k^2}  \right)
-\frac{(\partial \mu_p/\partial n_n+2D_{pn}k^2)^2}{\partial \mu_n/\partial n_n+2D_{nn}k^2}
-\frac{(4\pi e^2/ k^2)^2}{\partial \mu_e/\partial n_e+D_{ee}k^2+4\pi e^2/k^2}.
\label{Vdyn-1}
\end{eqnarray}
%\end{widetext}
The  chemical potential  $\mu_n$ and $\mu_p$ {\bf are} defined as
\begin{equation}
\mu_n=\left(\frac{\partial E_b}{\partial n_n}  \right)_{n_p}, \quad \mu_p=\left(\frac{\partial E_b}{\partial n_p}  \right)_{n_n},
\label{Def-chem-pot}
\end{equation}
where $n_n$ and $n_p$ the number densities of neutrons and protons respectively and $E_b$ the energy per baryon (including protons and neutrons).
It is worth to discuss here with more details the gradient terms $D_{ij}$ ($ i,j=p,n$). These terms are in general functions of the density but we treat them as constants. Moreover, since the models  used in the present work do not have gradient terms we fix them in an approximate way (see the discussion at the end of the subsection).
 Now, in  Eq.~(\ref{Vdyn-1}) neglecting the factor $D_{ee}$ and
retaining for consistency only terms of order of $k^2$ in the curvature term, due to the momentum wave-number taking small values, we find the well known approximation \cite{Baym-71}
\begin{equation}
U_{\rm dym}(k,n)=U_0(n)+\xi k^2+\frac{4\pi e^2}{k^2+k_{TF}^2},
\label{Vdyn-appr}
\end{equation}
where
\begin{equation}
U_0(n)=\frac{\partial \mu_p}{\partial n_p}-\frac{(\partial \mu_p/\partial n_n)^2}{\partial \mu_n/\partial n_n},
\label{V0}
\end{equation}
\begin{equation}
\xi=2(D_{pp}+2D_{np}\zeta+D_{nn}\zeta^2), \quad \zeta=-\frac{\partial \mu_p/\partial n_n}{\partial \mu_n/\partial n_n}
\label{zeta-1}
\end{equation}
and also
\begin{equation}
k^2_{TF}=\frac{4}{\pi}\frac{e^2}{\hbar c}k_e^2=\frac{4}{\pi}\frac{e^2}{\hbar c}\left(3\pi^2 x n\right)^{2/3}.
\label{Ktf-1}
\end{equation}
In Eq.~(\ref{Ktf-1}) $k_e$ is the electron Fermi momentum and $x=n_e/n$ is the electron fraction. In addition, the electron chemical potential $\mu_e$ is given by
\begin{equation}
\mu_e=\hbar c (3\pi^2 n_e)^{1/3}.
\label{mu-e-1}
\end{equation}
We note that here we have  $D_{ij}=B_{ij}/n_0$ according to the notation of Baym et al.~\cite{Baym-71}.
In the specific case where $D_{pp}=D_{nn}=D_{pn}/2$ we get
\[\xi=2D_{nn}(1+4\zeta +\zeta^2).  \]
The effective interaction $U_{\rm dym}(k,n)$, given by Eq.~(\ref{Vdyn-appr}), for a fixed value of the density $n$, has a minimum at $k=Q$ given by
\begin{equation}
Q^2=\sqrt{\frac{4\pi e^2}{ \xi}}-k_{TF}^2.
\label{Q-min}
\end{equation}
Now, replacing $k=Q$ in Eq.~(\ref{Vdyn-appr}) we find  the least stable modulation
\begin{equation}
U_{\rm dyn}(Q,n)=U_0(n)+4\sqrt{\pi \alpha \hbar c\xi}-4 \alpha \xi  \left(9\pi x^2 n^2 \right)^{1/3}, \quad \alpha=e^2/\hbar c.
\label{VQ-min}
\end{equation}
The transition density $n_{\rm t}$ is determined now from the condition $U_{\rm dyn}(Q,n_{\rm t})=0$.
The basic ingredients of Eq.~(\ref{VQ-min}) are the energy per baryon of nuclear matter $E_b$ (and consequently the chemical potentials of neutrons and protons) and also the proton fraction $x$. Now, it is important  to discuss the selected values of the gradient terms $D_{ij}$.
Following the formalism introduced by Bethe~\cite{Bethe-68} and elaborated by Ravenhall {\it et al.}~\cite{Ravenhall-72,Ravenhall-83} and Steiner {\it et al.}~\cite{Steiner-05}, we consider that the total energy density of semi-infinite matter is given by
\begin{equation}
{\cal E}_b(n)=nE_b(n,x=0.5)+D\left(\frac{dn(z)}{dz}  \right)^2,
\label{D-1}
\end{equation}
where $z$ is the distance of the surface and $D$ a constant related with the coefficients  $D_{ij}$ according to $D=3D_{nn}/2=3D_{pp}/2=3D_{np}/4$.
The quantity $D$ can be determined either from the surface energy of symmetric nuclear matter or from the surface thickness of symmetric nuclei~\cite{Lattimer-07}. By minimizing the total energy according to $\int_{-\infty}^{\infty}{\cal E}_b(n) dz$ with respect to the baryon density $n(z)$ and for fixed number of baryons we found (see also the recent work~\cite{Tews-017})
\begin{equation}
n\left(E_b(n,x=0.5)-\lambda\right)=D\left(\frac{dn(z)}{dz}  \right)^2,
\label{D-2}
\end{equation}
where $\lambda$ is the Lagrange multiplier  fixed by  the equation $\lambda=E_b(n_s,x=0.5)=E_0$ ($n_s$ is the saturation density of symmetric nuclear matter). We define the function
\begin{equation}
g(u)=u\left(\frac{E_b(n,x=0.5)-\lambda}{E_{\rm kin}}\right),
\label{fu-1}
\end{equation}
where $u=n/n_s$ and  $E_{\rm kin}$ is the kinetic energy at the saturation density $n_s$. Now, the surface thickness is written
\begin{equation}
t_{90-10}=\sqrt{\frac{Dn_s}{E_{\rm kin}}}\int_{0.1}^{0.9}\frac{1}{\sqrt{g(u)}}du.
\label{t-1}
\end{equation}
The surface tension of the  symmetric nuclear matter $\sigma_{\rm snm}$  defined as
\begin{eqnarray}
\sigma_{\rm snm}\equiv\int_{-\infty}^{+\infty}\left({\cal E}_b-\lambda n\right)dz
=2\int_{-\infty}^{+\infty}\left(nE_b(n,x=0.5)-\lambda n\right)dz
\label{sigma-2}
\end{eqnarray}
can be written also as
\begin{equation}
\sigma_{\rm snm}=2\sqrt{D E_{\rm kin} n_s^3}\int_{0}^{1}\sqrt{g(u)}du.
\label{sigma-2}
\end{equation}
The function $g(u)$ is defined for each applied nuclear model and the parameter $D$ is varied in an interval which leads to reasonable values for the surface thickness $t_{90-10}$ and the surface tension $\sigma_{\rm snm}$. In particular, the gradient terms related with  $t_{90-10}$  and $\sigma_{\rm snm}$ are selected in a such a way that these quantities are close to the empirical values~\cite{Xu-09,Tews-017,Hebeler-13,Lim-017,Mayers-69,Centelles-98,Kolehmainen-85,Togawa-09,Hofer-89}.
In Fig.~1 we plot, for the considered  models, the dependence of the surface tension and surface thickness on $D$.  We found that the value $D=72 \ {\rm MeV  \ fm^5}$ (and consequently $D_{nn}=D_{pp}=48 \ {\rm MeV  \ fm^5} $) leads to reasonable values both for  the surface thickness and surface tension. The results are presented also  in Table~1.  Of course one can fix the values of $D_{ij}$ for each model separately in order to keep the uniformity of the gradient term but we considered the present approximation to be  reasonable. In any case a more systematic study of the effects of the gradient term on the transition density  has been presented and discussed also in Refs.~\cite{Xu-09,Lim-017}.

It should be noted that neglecting in Eq.~(\ref{Vdyn-appr}) the gradient and the Coulomb contribution  (the second and third  term respectively),   the dynamical method  is reduced  to the thermodynamical one~\cite{Xu-09,Kubis-07a,Kubis-07b,Moustakidis-010,Moustakidis-012}. In this case, the solution of the equation $U_0(n_{\rm t})=0$ leads to the transition density $n_{\rm t}$. Obviously, the contribution of the gradient and the Coulomb term, to the estimation of $n_{\rm t}$ and $P_{\rm t}$, can be studied separately.

%%%%%%%%%%%%%%%%%%%%%%%%%%%%%%%%%
\subsection{Symmetry energy}
%%%%%%%%%%%%%%%%%%%%%%%%%%%%%%%%%%
The  symmetry energy plays an important role on the determination of the transition density and the corresponding pressure  and is  a key quantity  to explain in general many neutron star properties and dynamical processes~\cite{Epja-014}.  We consider that  the energy per particle of nuclear matter $E_b(n,I)$ can be expanded around the asymmetry parameter $I$ as~\cite{Moustakidis-012}
\begin{equation}
E_b(n,I)=E_b(n,I=0)+E_{\rm sym,2}(n)I^2+E_{\rm sym,4}(n)I^4+\cdots+E_{\rm sym,2k}(n)I^{2k}+\cdots
\label{esym-exp-1}
\end{equation}
where $I = (n_n-n_p)/n = 1-2x$ ($x$ is the proton
fraction $n_p/n$). The coefficients of the expansion~(\ref{esym-exp-1}) are given by the expression
\begin{equation}
E_{\rm sym,2k}(n)=\frac{1}{(2k)!}\frac{\partial^{2k}E_b(n,I)}{\partial I^{2k}}|_{I=0}.
\label{Esy-coef}
\end{equation}
The nuclear symmetry energy $E_{\rm sym}(n)$ is defined as the coefficient of the quadratic term, that is
\begin{equation}
E_{\rm sym}(n)\equiv E_{\rm sym,2}(n)=\frac{1}{2!}\frac{\partial^{2}E_b(n,I)}{\partial I^{2}}|_{I=0}
\label{Esy-def}
\end{equation}
and the slope of the symmetry energy $L$ at the nuclear saturation density $n_s$, which is an indicator of the stiffness of the EoS,  is defined as
\begin{equation}
L=3n_s\frac{dE_{\rm sym}(n)}{d n}|_{n=n_s}.
\label{L-def}
\end{equation}
In the framework of the parabolic approximation (PA) the energy per particle is given by the expression
\begin{equation}
E_b(n,x)\simeq E_b\left(n,I=0\right)+I^2E_{sym}^{PA}(n),
\label{PA-1}
\end{equation}
where $E_{\rm sym}^{PA}(n)$ is simply defined as
\begin{equation}
E_{\rm sym}^{PA}(n)=E_b(n,I=1)-E_b\left(n,I=0\right).
\label{PA-2}
\end{equation}
In $\beta$-stable nuclear matter the following processes take place simultaneously
\begin{equation}
n\rightarrow p+e^{-}+\bar{\nu}_e, \quad p+e^{-} \rightarrow n+{\nu}_e
\label{beta-1}
\end{equation}
and considering that neutrinos generated in these reactions have left the system, the chemical equilibrium condition takes the form
\begin{equation}
\mu_n=\mu_p+\mu_e.
\label{beta-1}
\end{equation}
It is easy to show that after some algebra we get~\cite{Moustakidis-015} (see also the Appendix)
\begin{equation}
\mu_n-\mu_p=\left(-\frac{\partial E_b}{\partial x}\right)_n.
\label{beta-2}
\end{equation}
Finally, using also Eq.~(\ref{mu-e-1}), we found
\begin{equation}
\left(\frac{\partial E_b}{\partial x}\right)_n=-\hbar c (3\pi^2 x n)^{1/3}.
\label{beta-3}
\end{equation}
Equation~(\ref{beta-3}) is the most general relation that
determines the proton fraction of $\beta$-stable matter and we will mention it  hereafter  as a full expansion (FE).
Now the total energy per particle of  neutron star matter $E(n,x)$ will be given by the sum of the energy per baryon and electron energy, that is
\begin{equation}
E(n,x)=E_b(n,x)+E_e(n,x),
\label{E-total}
\end{equation}
where the fraction $x$ is determined, in general, by Eq.(\ref{beta-3}). The electrons are considered as a non-interacting Fermi gas and consequently~\cite{Shapiro-83}
\begin{equation}
E_e(n,x)=\frac{3}{4}\hbar c \left(3\pi^2 x^4n^4\right)^{1/3}.
\label{Ee-1}
\end{equation}
Accordingly, the total pressure is decomposed also
into baryon and lepton contributions
\begin{equation}
P(n,x)=P_b(n,x)+P_e(n,x), \label{P-all-1}
\end{equation}
where by definition
\begin{equation}
P_b(n,x)=n^2\left(\frac{\partial E_b}{\partial  n}\right)_x. \label{Pb-1}
\end{equation}
The contribution of the electrons  to the total pressure is equal to
\begin{equation}
P_e(n,x)=\frac{1}{12\pi^2}\frac{\mu_e^4}{(\hbar c)^3}=\frac{\hbar
c}{12 \pi^2}\left(3\pi^2 xn\right)^{4/3}. \label{Pe-2}
\end{equation}
Now, the transition pressure $P_{\rm t}$  in the case  of  the FE, is given  by
the equation
\begin{equation}
P_{\rm t}^{FE}(n_{\rm t},x_{\rm t})=n_{\rm t}^2\left(\frac{\partial E_b}{\partial
n}\right)_{n=n_{\rm t}}+\frac{\hbar c}{12 \pi^2}\left(3\pi^2
x_{\rm t}n_{\rm t}\right)^{4/3}. \label{Pr-tra}
\end{equation}
In  the case of the parabolic approximation,  the use of  Eq.~(\ref{beta-3}) with the definition~(\ref{PA-1}) leads to the determination of the proton fraction by the equation
\begin{equation}
4(1-2x)E_{\rm sym}^{PA}(n)=\hbar c(3 \pi^2 n
x)^{1/3}.
\label{b-equil-2}
\end{equation}
In this case the transition pressure  $P_{\rm t}^{PA}$ is given by the relation~\cite{Moustakidis-012}
%\begin{widetext}
\begin{eqnarray}
P_{\rm t}^{PA}(n_{\rm t},x_{\rm t})=n_{\rm t}^2 \left[ \left(\frac{{\rm d}
E_b(n,x=0.5)}{{\rm d} n}\right)_{n=n_{\rm t}}
+\left(\frac{{\rm d}
 E_{\rm sym}^{PA}(n)}{{\rm d} n}\right)_{n=n_{\rm t}}(1-2x_{\rm t})^2 \right]
+\frac{\hbar c}{12 \pi^2}\left(3\pi^2 x_{\rm t}n_{\rm t}\right)^{4/3}.
\label{Pr-tra-para}
\end{eqnarray}
%\end{widetext}

%%%%%%%%%%%%%%%%%%%%%%%%%%%%%%%%
%\subsection{The coefficients of the curvature term}
%%%%%%%%%%%%%%%%%%%%%%%%%%%%%%%%%
%It is worth to discuss the accuracy of the selected values of $D_{ij}$.

%%%%%%%%%%%%%%%%%%%%%%%%
\section{The Models}
%%%%%%%%%%%%%%%%%%%%%%%%%%%%%%%
In the present work we employed various nuclear models, which are suitable for reproducing
the bulk properties of nuclear matter at low densities, close
to saturation density as well as the maximum observational
neutron star mass. In particular, in each case,
the energy per particle of nuclear matter $E_b(n,I)$ is given as a
function of the baryonic number density n and the asymmetry
parameter $I$ (or the proton fraction $x$).
%%%%%%%%%%%%%%%%%%%%%%%%%%%%%%%%%
\subsection{MDI model}
%%%%%%%%%%%%%%%%%%%%%%%%%%%%%%%%%%
The momentum-dependent interaction (MDI) model used here, was already presented and analyzed in previous papers~\cite{Prakash-97,Moustakidis-08}. The MDI model is designed to reproduce the results of the microscopic calculations of both nuclear and neutron rich
matter at zero temperature and it can be extended to finite temperature. The energy per baryon  at $T=0$, is given by
%\begin{widetext}
\begin{eqnarray}
E_b(n,I)&=&\frac{3}{10}E_F^0u^{2/3}\left[(1+I)^{5/3}+(1-I)^{5/3}\right] +
\frac{1}{3}A\left[\frac{3}{2}-(\frac{1}{2}+x_0)I^2\right]u
+
\frac{\frac{2}{3}B\left[\frac{3}{2}-(\frac{1}{2}+x_3)I^2\right]u^{\sigma}}
{1+\frac{2}{3}B'\left[\frac{3}{2}-(\frac{1}{2}+x_3)I^2\right]u^{\sigma-1}}\nonumber
\\
&+&\frac{3}{2}\sum_{i=1,2}\left[C_i+\frac{C_i-8Z_i}{5}I\right]\left(\frac{\Lambda_i}{k_F^0}\right)^3
\left(\frac{\left((1+I)u\right)^{1/3}}{\frac{\Lambda_i}{k_F^0}}-
\tan^{-1} \frac{\left((1+
I)u\right)^{1/3}}{\frac{\Lambda_i}{k_F^0}}\right)\nonumber \\
&+&
\frac{3}{2}\sum_{i=1,2}\left[C_i-\frac{C_i-8Z_i}{5}I\right]\left(\frac{\Lambda_i}{k_F^0}\right)^3
\left(\frac{\left((1-I)u\right)^{1/3}}{\frac{\Lambda_i}{k_F^0}}-
\tan^{-1}
\frac{\left((1-I)u\right)^{1/3}}{\frac{\Lambda_i}{k_F^0}}\right).
 \label{e-T0}
\end{eqnarray}
%\end{widetext}
%
In Eq.~(\ref{e-T0}) the ratio $u$ is defined as   $u=n/n_s$, with $n_s$ denoting the
equilibrium symmetric nuclear matter density (or saturation density), $n_s=0.16$ fm$^{-3}$.
The parameters $A$, $B$, $\sigma$, $C_1$, $C_2$ and
$B'$ which appear in the description of symmetric nuclear matter take the values $A=-46.65$, $B=39.45$, $\sigma=1.663$,
$C_1=-83.84$, $C_2=23$ and $B'=0.3$. They are determined  by the requirement that Eq.~(\ref{e-T0}) reproduces the binding energy  $E_b(n=n_s,I=0)=-16$ {\rm MeV}
at the saturation density $n_s=0.16\  fm^{-3}$ and the incompressibility is $K=240$ {\rm MeV}.
The finite range parameters are $\Lambda_1=1.5 k_F^{0}$ and $\Lambda_2=3 k_F^{0}$ with $k_F^0$
being the Fermi momentum at the saturation density $n_s$. By suitably choosing the parameters $x_0$, $x_3$, $Z_1$, and $Z_2$, it is
possible to obtain different form for the density dependence of the symmetry energy as well as for the value of the
slope parameter L and the value of the symmetry energy at the saturation density~\cite{Moustakidis-015,Moustakidis-08}. Actually, for each value of L the
density dependence of the symmetry energy is adjusted so that the energy of pure neutron matter is comparable with those of
existing state-of-the-art calculations~\cite{Moustakidis-015,Moustakidis-08}.

%%%%%%%%%%%%%%%%%%%%%%%%%%
\subsection{Skyrme  model}
%%%%%%%%%%%%%%%%%%%%%%%%%%
The Skyrme functional providing the energy per baryon of asymmetric nuclear matter is given by the formula
\cite{Chabanat-97,Farine-97}
%\begin{widetext}
\begin{eqnarray}
E_b(n,I)&=&\frac{3}{10}\frac{\hbar
^2c^2}{m}\left(\frac{3\pi^2}{2}\right)^{2/3}n^{2/3}F_{5/3}(I)
+\frac{1}{8}t_0n\left[2(x_0+2)-(2x_0+1)F_2(I)\right]\nonumber\\
&+&\frac{1}{48}t_3n^{\sigma+1}\left[2(x_3+2)-(2x_3+1)F_2(I)\right] \\
&+& \frac{3}{40}\left(\frac{3\pi^2}{2}\right)^{2/3}n^{5/3}
\left[\frac{}{}\left(t_1(x_1+2)+t_2(x_2+2)\right)F_{5/3}(I)
+\frac{1}{2}\left(t_2(2x_2+1)-t_1(2x_1+1)\right)F_{8/3}(I)\right],
\nonumber
\end{eqnarray}
%\end{widetext}
%
where $\displaystyle F_m(I)=\frac{1}{2}\left[(1+I)^m+(1-I)^m
\right] $ and the parametrization is given in
Refs~\cite{Chabanat-97,Farine-97}.

%%%%%%%%%%%%%%%%%%%%%%%%%%%%%%%%%%%%%%
\subsection{The HLPS model}
%%%%%%%%%%%%%%%%%%%%%%%%%%%%%%%%%%%%%%%
Recently, Hebeler {\it et al.}~\cite{Hebeler-010,Hebeler-13} performed microscopic calculations based on chiral effective field theory interactions  to constrain the properties of neutron-rich matter below nuclear densities. It explains the massive neutron stars of $M=2 M_{\odot}$.  In this model the energy per particle is given by~\cite{Hebeler-13} (hereafter HLPS model)
%\begin{widetext}
\begin{eqnarray}
E_b(u,x)&=&\frac{3T_0}{5}\left(x^{5/3}+(1-x)^{5/3}  \right)(2u)^{2/3}
-T_0\left[(2\alpha-4\alpha_L)x(1-x)+\alpha_L\right]u \nonumber \\
&+&
T_0\left[(2\eta-4\eta_L)x(1-x)+\eta_L\right]u^{\gamma},
\label{ED-HLPS}
\end{eqnarray}
%\end{widetext}
%
where $T_0=(3\pi^2n_0/2)^{2/3}\hbar^2/(2m)=36.84$ MeV.
The parameters $\alpha$, $\eta$, $\alpha_L$ and $\eta_L$ are determined by combining   the saturation properties of symmetric nuclear matter and the microscopic calculations for neutron matter~\cite{Hebeler-010,Hebeler-13}. The parameter $\gamma$ is used to adjust the values of the incompressibility $K$ and influences the range of the values of the symmetry energy and its density derivative. In the present work we employ the values $\gamma=4/3$,  $\alpha=5.87$, $\eta=3.81$,  also $\alpha_L=1.3631$ with $\eta_L=0.7596$ (soft and intermediate equation of state) and $\alpha_L=1.53148$ with $\eta_L=1.02084$ (stiff equation of state)~\cite{Hebeler-13}.

%%%%%%%%%%%%%%%%%%%%%%%%%%%%%%%%%%%%%%%%%%%%
\section{Applications}
%%%%%%%%%%%%%%%%%%%%%%%%%%%%%%%%%%%%%%%%%%%%%%%%
In the following we provide some applications of the crust-core interface in astrophysics.  Firstly, we provide a derivation of  model-independent relations between the mass of the crust and the transition pressure and also one between the latter  and the slope of the symmetry energy. Secondly,  we concentrate our study on the effects of the transition density and transition pressure on  a)   the oscillation frequencies obtained from observations of quasi-periodic oscillations (QPOs), b)  the thermal relaxation time of the crust during the cooling process of a hot neutron star, c)  the crustal fraction of the moment of inertia and its effects on the creation of neutron star glitches and d)  the conditions for  the r-mode instabilities of rotating neutron stars.
%%%%%%%%%%%%%%%%%%%%%
\subsection{Radius and mass of the crust}
%%%%%%%%%%%%%%%%%%%%%%%%%%%%555
The radius $R_{\rm crust}$ and the mass $M_{\rm crust}$ of the crust play an important role in various neutron star properties as we will present below. In addition it will be useful and instructive to find analytical approximations to relate the above quantities both with the bulk neutron star properties as well as, if it is possible, with some details of the neutron star EoS. The starting point of  this effort is the well known
Tolman-Oppenheimer-Volkoff (TOV) equations~\cite{Tolman-39,Oppenheimer-39} which describe the structure of a neutron star and have the form
\begin{equation}
\frac{dP(r)}{dr}=-\frac{G{\cal E}(r) M(r)}{c^2r^2}\left(1+\frac{P(r)}{{\cal E}(r)}\right)\left(1+\frac{4\pi P(r) r^3}{M(r)c^2}\right) \left(1-\frac{2GM(r)}{c^2r}\right)^{-1},
\label{TOV-1}
\end{equation}
%%%%%%
\begin{equation}
\frac{dM(r)}{dr}=\frac{4\pi r^2}{c^2}{\cal E}(r).
\label{TOV-2}
\end{equation}
Recently,  Zdunik {\it et al.}~\cite{Zdunik-017}, starting from the assumption that  the term $4\pi P(r) r^3/M(r)c^2$ is very small compared to $1$
and employing also the relation
\begin{equation}
\frac{dP}{{\cal E}+P}=\frac{d\mu}{\mu}
\label{dP-1}
\end{equation}
where
\begin{equation}
\mu=\frac{{\cal E}+P}{n}
\label{chem-1}
\end{equation}
is the baryon chemical potential,  found that the radius of such a star and the corresponding values for its crust and core
are given respectively   by the expressions
\begin{equation}
R=\frac{R_{\rm core}}{1-(h_{\rm t}-1)(R_{\rm core}c^2/2GM-1)},
\label{Cradii-1}
\end{equation}
\begin{equation}
\frac{R_{\rm crust}}{R}=\frac{(h_{\rm t}-1)(1-2\beta)}{h_{\rm t}-1+2\beta}
\label{Cradii-2}
\end{equation}
and
\begin{equation}
\frac{R_{\rm core}}{R}=\frac{2\beta h_{\rm t}}{h_{\rm t}-1+2\beta}.
\label{Cradii-3}
\end{equation}
In Eqs~(\ref{Cradii-1}), (\ref{Cradii-2}) and (\ref{Cradii-3}) $\beta=GM/Rc^2$ is the compactness parameter and $h_{\rm t}$ is defined as
\begin{equation}
h_{\rm t}=\left(\frac{\mu_{\rm t}}{\mu_0}\right)^2,
\label{h-1}
\end{equation}
where $\mu_{\rm t}$ and  $\mu_0$  are the chemical potentials at the crust-core interface and on the surface respectively. Actually, at the transition density we have  ${\cal E}_{\rm t}\gg P_{\rm t}$ and consequently  the above relation becomes
\begin{equation}
h_{\rm t}\simeq\frac{1}{\mu_0^2}\left(\frac{{\cal E}_{\rm t}}{n_{\rm t}}\right)^2.
\label{ht-aprox}
\end{equation}
In the present work  we consider that  $\mu_0=930.4$ MeV~\cite{Haensel-07}. According to Eqs.~(\ref{Cradii-1}) and (\ref{Cradii-2}) the effect of the  EoS is included indirectly via the compactness parameter $\beta$ and the radius $R$ and directly via the factor $h_{\rm t}$ which is related with the energy per particle of neutron star matter at the transition density.
It is worth to point out  that a similar expression  has been found by Lattimer et al.~\cite{Lattimer-07} by just replacing the quantity  $h_{\rm t}$ by  ${\cal H}$
where
\[ {\cal H}={\rm e}^{2(\mu_{\rm t}-\mu_0)/m_bc^2}. \]
Obviously for  $(\mu_{\rm t}/\mu_0)^2-1\ll 1$   the two approximations coincide. In the present work we will employ the approximations (\ref{Cradii-2}) and (\ref{Cradii-3}).

Now,  we will  derive  an  approximate expression   for the $M_{\rm crust}$ in comparison with recent studies~\cite{Zdunik-017}. Firstly,  we neglect the term $4\pi P(r) r^3/M(r)c^2$ in the first of the TOV equations, which is three orders of  magnitude  less than unity {\bf in} the region from the crust-core interface to the surface. We  consider  also the approximation $r\simeq R_{\rm core}$ which introduces an error at most 10$\%$  (which appears just close to the surface) and mainly for low values of neutron star mass. The combination of the TOV equations now leads to the equation
\begin{equation}
\frac{dP(r)}{dM(r)}=-\frac{G M(r)}{4\pi R_{\rm core}^4(1-2GM(r)/R_{\rm core}c^2)}
\label{Mcr-ap-1}
\end{equation}
and integrating from the crust-core edge to the surface we get
\begin{equation}
\int_{P_{\rm t}}^{0} dP=-\frac{c^4}{4\pi R_{\rm core}^2 G}\int_{x_{\rm t}}^{x_{\rm s}} \frac{x}{1-2x}dx,
\qquad x\equiv x(r)=\frac{G M(r)}{R_{\rm core}c^2}.
\label{aprox-1}
\end{equation}
The analytical value of the integral is
\begin{equation}
\int_{x_{\rm t}}^{x_{\rm s}} \frac{x}{1-2x}dx=\frac{1}{4}\left[ 2(x_{\rm t}-x_{\rm s})+\ln \left(\frac{1-2x_{\rm t}}{1-2x_{\rm s}}\right)\right],
\qquad x_{\rm t}=\frac{GM_{\rm core}}{R_{\rm core}c^2}, \ x_{\rm s}=\frac{GM}{R_{\rm core}c^2}.
\label{aprox-2}
\end{equation}
After some algebra we get
%\begin{equation}
%P_{\rm t}=\frac{c^4}{16\pi R_{\rm core}^2 G}\left[-\frac{2GM_{\rm crust}}{R_{\rm core}c^2}-\ln\left({1-\frac{2GM_{\rm crust}/R_{\rm core}c^2}{1-2GM_{\rm %core}/R_{\rm core}c^2}}  \right)\right]
%\label{aprox-3}
%\end{equation}
%
%or
\begin{equation}
P_{\rm t}=\frac{c^4}{16\pi R_{\rm core}^2 G}\left[-\frac{2M_{\rm crust}\beta_{\rm core}}{M_{\rm core}}-\ln\left(1-\frac{2M_{\rm crust}\beta_{\rm core}/M_{\rm core}}{1-2\beta_{\rm core}}  \right)\right], \quad \beta_{\rm core}=\frac{GM_{\rm core}}{R_{\rm core}c^2}.
\label{aprox-3-new}
\end{equation}
The above approximation relates  the microscopic quantity $P_{\rm t}$ with the macroscopic quantities $M_{\rm core}, M_{\rm crust}, R_{\rm core}$ and consequently only indirectly depends on the EoS. The observational determination of the crustal and core mass as well as the core radius will impose constraints on the values of $P_{\rm t}$ and consequently on  the EoS and subsaturation densities.   Now, in order to proceed further and considering that
\[a=\frac{2M_{\rm crust}\beta_{\rm core}/M_{\rm core}}{1-2\beta_{\rm core}}\ll 1  \]
we employ the approximation
\begin{equation}
\ln(1-a)=-a-\frac{a^2}{2}+{\cal O}(a^3).
\nonumber
\label{exp-lna}
\end{equation}
In this case,  the transition pressure is approximated by the expression
\begin{equation}
P_{\rm t}=\frac{G M_{\rm crust}M_{\rm core}}{4\pi R_{\rm core}^4 (1-2\beta_{\rm core})}\left(1+\frac{M_{\rm crust}/M_{\rm core}}{2(1-2\beta_{\rm core})}\right),
\label{2st-aprox}
\end{equation}
and therefore
\begin{equation}
M_{\rm crust}=M_{\rm core}\left( 1-\frac{2GM_{\rm core}}{R_{\rm core}c^2}\right)\left(\sqrt{\frac{8\pi R_{\rm core}^4P_{\rm t}}{GM_{\rm core}^2 }+1  }-1\right).
\label{3st-aprox}
\end{equation}
Considering also that  $$\frac{8\pi R_{\rm core}^4P_{\rm t}}{GM_{\rm core}^2}  \ll 1$$ we get
\begin{equation}
\sqrt{\frac{8\pi R_{\rm core}^4P_{\rm t}}{GM_{\rm core}^2 }+1  }\simeq 1+\frac{1}{2}\frac{8\pi R_{\rm core}^4P_{\rm t}}{GM_{\rm core}^2}
\label{Exp-1}
\end{equation}
and finally  we  find
\begin{equation}
M_{\rm crust}=\frac{4\pi P_{\rm t}R_{\rm core}^4}{GM_{\rm core}}\left( 1-\frac{2GM_{\rm core}}{R_{\rm core}c^2}\right).
\label{1st-aprox}
\end{equation}
Actually, Eq.~(\ref{1st-aprox}) has been  provided by Zdunik {\it et al.}~\cite{Zdunik-017}. It is right to point out that a similar expression has been derived by Pethick and Ravenhall~\cite{Petchik-95a}. In particular they provided the approximation
\[M_{\rm crust}\simeq \frac{4\pi P_{\rm t}R^4}{GM}\left( 1-\frac{2GM}{Rc^2}\right).  \]
Recently, Baym {\it et al.}~\cite{Baym-017} following a similar approach,  provided the approximation
\[ M_{\rm crust}\simeq \frac{2\pi P_{\rm t}R_{\rm core}^4}{GM_{\rm core}}\left( 1-\frac{2GM_{\rm core}}{R_{\rm core}c^2}\right). \]
which is  half of the value obtained by   Zdunik {\it et al.}~\cite{Zdunik-017}.  It is also worth to mention the approximation found in Ref.~\cite{Fattoyev-10} where the authors following the same assumptions for the the solid crust but considering a specific equation of state i.e. the well known polytropic one $P({\cal E})=K{\cal E}^{4/3}$, obtained  very simple analytical expressions for the crustal moment of inertia and mass. In their work the crustal mass is given by~\cite{Fattoyev-10}
\begin{equation}
M_{\rm cr}\approx 8\pi R_{\rm core}^3 P_{\rm t}\left(\frac{R_{\rm core}}{R_s}-1\right)\left[1+\frac{32}{5}\left(\frac{R_{\rm core}}{R_s}-\frac{3}{4}  \right)\frac{P_{\rm t}}{{\cal E}_{\rm t}}+\dots  \right]
\label{Piek-Mcr}
\end{equation}
where $R_s=2GM/c^2$. Obviously, the leading order terms of the  approximations (\ref{3st-aprox}) and (\ref{Piek-Mcr}) coincide.

Now we can proceed further by considering the accurate approximation
\begin{equation}
P_{\rm t}=\frac{G M_{\rm crust}M_{\rm core}}{4\pi R_{\rm core}^4 (1-2\beta_{\rm core})},
\label{2st-aprox-1}
\end{equation}
and also the empirical assumptions $M_{\rm crust}\simeq (0.02-0.03) M_{\odot}$, $M_{\rm core}=M$, $R_{\rm core}=0.9 R$ which hold for a neutron star with mass $M=1.4 \ M_{\odot}$.
In this case, considering also that the corresponding radius lies in the interval $11 \ {\rm km} \leq R_{1.4} \leq 14 \ {\rm km}$, we find the semi-analytical relation
\begin{equation}
P_{\rm t}=\left(\frac{C_{\rm t}(1.4M_{\odot})}{R_{1.4}}\right)^4 \ {\rm MeV}\cdot {\rm fm}^{-3},
\label{Pt-anal-1}
\end{equation}
where
\[C_{\rm t}(1.4M_{\odot})=10.25\pm0.71 \ {\rm km}.  \]
The higher the values of the observational measure of $R_{1.4}$, the higher the accuracy for determination of $P_{\rm t}$. Moreover, the combination of  relation (\ref{Pt-anal-1}) with the empirical prediction of Lattimer and Prakash~\cite{Lattimer-01}
\begin{equation}
P(n_s)=\left(\frac{R_{1.4}}{C_s(n_s,1.4M_{\odot})}\right)^4 \ {\rm MeV}\cdot {\rm fm}^{-3}, \ C_s(n_s,1.4M_{\odot})=9.52\pm0.49 \ {\rm km}
\label{Lat-pr}
\end{equation}
where $P(n_s)$ is the pressure of neutron star matter at the saturation density,  helps to constrain the EoS at subsaturation densities. Considering also  that at the saturation density $n_s$, in a good approximation,  the pressure is given by~\cite{Lattimer-07}
\begin{equation}
P(n_s)=n_s^2\left(\frac{dE_{\rm sym}(n)}{dn}\right)_{n=n_s}(1-2x)^2+n_sx(1-2x)E_{\rm sym}(n_s)
\label{Pns-aapr-full}
\end{equation}
where the proton fraction is  $x\simeq(4E_{\rm sym}(n_s)/\hbar c)^3/(3\pi^2n_s)$ and also  $E_{\rm sym}(n_s)\simeq 30$ MeV. Then, after some algebra we find the expression
\begin{equation}
P(n_s)\simeq \left(\frac{n_sL}{3}C_s\right)\ {\rm MeV}\cdot {\rm fm}^{-3}, \quad  C_s=0.90\pm0.05.
\label{Ps-L}
\end{equation}
Finally, combining Eqs.~(\ref{Pt-anal-1}), (\ref{Lat-pr}) and (\ref{Ps-L}) by eliminating the radius $R_{1.4}$ and taking into account that $n_s\simeq 0.16\ {\rm fm}^{-3}$ we find that
\begin{equation}
P_{\rm t}=\left(\frac{C_L}{L}\right)\ {\rm MeV}\cdot {\rm fm}^{-3}, \quad C_L=32.08\pm 15.80\ {\rm MeV},
\label{Pt-L}
\end{equation}
where $L$ is given in MeV. It is worthwhile to notice that Eq.~(\ref{Pt-L}) has been constructed in a model independent way by using only the TOV equations and the empirical formulae~(\ref{Lat-pr}).  We would like to emphasize here that Eq.~(\ref{Pt-L}) has been obtained  considering that the value of the symmetry energy at the saturation density is constant i.e. $E_{\rm sym}(n_s)= 30$ MeV. However, in the case where both $L$ and $E_{\rm sym}(n_s)$  vary,  the dependence of  $P_{\rm t}$ {\bf on} $L$ may exhibit a different behavior as found for example in Ref.~\cite{Newton-13}.  According to Eq.~(\ref{Pt-L}) the  stiffness of the EoS acts against the solidification of nuclear matter providing  theoretical agreement with   and interpretation   of previous results~\cite{Xu-09,Moustakidis-015,Boquera-17,Wei-018}. Although the uncertainty in Eq.~(\ref{Pt-L}) is relatively high, the exhibited $P_{\rm t}-L$ dependence is qualitatively correct.

%%%%%%%%%%%%%%%%%%%%%%%%%%%%%%%%
%\subsection{Application I: Radius and mass of the crust }
%%%%%%%%%%%%%%%%%%%%%%%%%%%%%%%%%%

%%%%%%%%%%%%%%%%%%%%%%%%%%%%%%%%
\subsection{Neutron star  oscillation frequencies }
%%%%%%%%%%%%%%%%%%%%%%%%%%%%%%%%%%
Information about radii can be obtained also from observations of quasi-periodic oscillations in the X-ray emissions  caused most likely by the  torsional vibration of the crust of a neutron star (for more details see the discussion of Lattimer {\it et al.}~\cite{Lattimer-07}). Now, considering the approximation $v_r\simeq u_t$ (where $v_r$ and $v_t$ are the average radial and transverse shear speed respectively) the authors in Ref.~\cite{Samuelsson-07} found  simple relations for the frequencies. In particular, the frequencies of the fundamental and higher modes can be written~\cite{Lattimer-07}
\begin{equation}
f_{n=0, l=2} \simeq 263.3 \left(\frac{{\rm km}}{R}\right)\sqrt{\frac{(h_{\rm t}-1+2\beta)(1-2\beta)}{\beta h_{\rm t}}}\  {\rm Hz},
\label{freq-1}
\end{equation}
\begin{equation}
f_{n>0} \simeq 1170 n  \left(\frac{{\rm km}}{R}\right)\frac{h_{\rm t}-1+2\beta}{h_{\rm t}-1} \  {\rm Hz}.
\label{freq-2}
\end{equation}
Obviously the measurement  of more than one of the frequencies can be used to identify $R$ and $\beta$  as functions of the quantity $h_{\rm t}$~\cite{Lattimer-07}. Moreover, eliminating $R$ from Eqs.~(\ref{freq-1}) and (\ref{freq-2}) a dependence $\beta\equiv \beta(h_{\rm t})$ can be found.
%%%%%%%%%%%%%%%%%%%%%%%%%%%%%%%%
\subsection{Thermal relaxation time  of the crust }
%%%%%%%%%%%%%%%%%%%%%%%%%%%%%%%%%%
The cooling of the core of a proto-neutron star, according to the accepted theory, is due to the neutrino emission. During the cooling process the star is not in thermal equilibrium as a consequence of the long thermal relaxation time of the crust. It is expected that the relaxation time is  of the  order 10-100 years~\cite{Lattimer-07}. After this time the surface comes into thermal equilibrium with the core. Actually, this is related to the specific heat and thermal conductivity of the crust as well as the crust radius. It was found that $t_w$ is given by the simple expression~\cite{Lattimer-07,Gnedin-01}
\begin{equation}
t_{\rm w}=\alpha t_1 \ ({\rm years)}, \quad  \alpha \equiv \left( \frac{R_{\rm crust}}{\rm km}\right)^2 \left(1-2MG/Rc^2\right)^{-3/2}
\label{Trel-1}
\end{equation}
where $t_1$ is the normalized relaxation time which depends solely on the macroscopic properties of nuclear matter including thermal conductivity and heat capacity~\cite{Gnedin-01}. For example for  non-superfluid stars and considering that the transition density is $n_{\rm t}=0.5n_0=0.08\ {\rm  fm}^{-3}$,
Gnedin {\it et al.}~\cite{Gnedin-01} suggested the values $t_1=28\pm 0.2 \ {\rm yr}$ for the rapidly cooling model and $t_1=22.9\pm 1.2 \ {\rm yr}$ for the slowly cooling models. Actually the effects of the crust-core interface are introduced through the value of the radius of the crust. Obviously, as already stated in Ref.~\cite{Lattimer-94} if the crust radius can be connected with the bulk neutron star properties $M$ and $R$ , then useful information concerning the neutron star structure can be inferred  from the observation of the surface cooling.

%%%%%%%%%%%%%%%%%%%%%%%%%%%%%%%%
\subsection{Crustal fraction of the moment of inertia and pulsar glitches}
%%%%%%%%%%%%%%%%%%%%%%%%%%%%%%%%%%
The pulsar glitches are sudden discontinuities in the spin-down of pulsars (for a recent review see Ref.~\cite{Haskell-015}). According to the more possible scenario they are due to the transfer of angular momentum from the superfluid component to the non-superfluid part of the crust~\cite{Ho-015}. Link {\it et al.}~\cite{Link-99} showed
that glitches represent a self-regulating instability for which
the star prepares over a waiting time. For example in the case of Vela pulsar  the observational glitches indicate that the moment of inertia of the crust must be at least 1.4 $\%$ of the total moment of inertia (although there are also some other explanations).
So, if glitches originate in the liquid of the inner crust, this
means that $ I_{\rm crust}/I>0.014$.

The crustal fraction of the moment of inertia $I_{\rm crust}/I$ can be expressed as a function of the total mass $M$  and radius  $R$
 with the only dependence on the equation of state arising from the values of $P_{\rm t}$ and $n_{\rm t}$. Actually, the major
dependence is on the value of $P_{\rm t}$, since $n_{\rm t}$ enters only as a
correction according to the following approximate
formula~\cite{Link-99}
%\begin{widetext}
\begin{eqnarray}
\frac{I_{\rm crust}}{I}&\simeq& \frac{28\pi P_{\rm t}
R^3}{3Mc^2}\frac{(1-1.67\beta-0.6\beta^2)}{\beta}\left(1+\frac{2P_{\rm t}}{n_{\rm t}mc^2}\frac{(1+5\beta-14\beta^2)}{\beta^2}
\right)^{-1}.  \label{inertia-1}
\end{eqnarray}
%\end{widetext}
%
The crustal fraction of the moment of inertia is particularly interesting since  it can be inferred from
observations of pulsar glitches, the occasional disruptions of the otherwise extremely regular pulsations from magnetized, rotating
neutron stars~\cite{Xu-09}.
More recently the authors in Ref.~\cite{Andersson-012,Chamel-012}, considering the  entrainment of superfluid neutrons in the crust,  found that the lower limit of $I_{\rm crust}/I$ must be larger than  $0.07$, in order to explain glitches. Moreover, Link~\cite{Link-014} who discussed in more detail the origin and the connection of the  moment of inertia of the crust and the core  concluded that  low values of $I_{\rm crust}/I$ must be expected. Very recently the authors in  Ref.~\cite{Watanabe-017} came to the conclusion   that the moment of inertia of the neutron superfluid in the crust is large enough {\bf so} that glitch models based on the superfluid neutrons in the inner crust cannot be ruled out. The above brief discussion reveals the necessity  of  further observational and theoretical work  in order to solve the  problem  of  glitches.
In any case,  it will be of interest to explore the effects of the transition density and pressure  on  $I_{\rm crust}/I$ compared to both the dynamical and thermodynamical method. Since the ratio $I_{\rm crust}/I$ is sensitive to  $n_{\rm t}$ and mainly {\bf to}  $P_{\rm t}$  useful constraints for the EoS close to the crust-core interface will be obtained from  future observation data from  pulsar glitches.

%%%%%%%%%%%%%%%%%%%%%%%%%%%%%%%%
\subsection{R-mode instability of a rotating neutron star}
%%%%%%%%%%%%%%%%%%%%%%%%%%%%%%%%%%
The r-modes are oscillations of rotating stars whose restoring
force is the Coriolis force
\cite{Lidblom-2000,Andersson-1998,Friedman-98,Friedman-99,Andersson-2001,Andersson-2003,Kokkotas-99,Andersson-99}.
The gravitational radiation-driven instability of these modes has
been proposed as an explanation for the observed relatively low
spin frequencies of young neutron stars and of accreting neutron
stars in low-mass X-ray binaries as well. This instability can
only occur when the gravitational-radiation driving time scale of
the r-mode is shorter than the time scales of the various
dissipation mechanisms that may occur in the interior of the
neutron star.

The nuclear EOS affects the time scales associated with the
r-mode, in two different ways. Firstly, EOS defines the radial
dependence of the mass density distribution $\rho(r)$, which is
the basic ingredient of the relevant integrals. Secondly, it
specifies  the core-crust transition density $\rho_{\rm t}$ and also the
core radius $R_{\rm core}$ which is the upper limit of the mentioned
integrals.

The critical angular velocity $\Omega_{\rm c}$, above which the {\it
r}-mode is unstable (for $m=2$) is given  by~\cite{Lidblom-2000}
\begin{equation}
\frac{\Omega_{\rm c}}{\Omega_0}=\left(-\frac{\tilde{\tau}_{GR}}{\tilde{\tau}_v}
\right)^{2/11}\left(\frac{10^8 \ K}{T}  \right)^{2/11},
\label{Omega-c-1}
\end{equation}
%%%%%
where $\Omega_0=\sqrt{\pi G\overline{\rho}}$,
$\overline{\rho}=3M/4\pi R^3$ is the mean density of the star, $T$ is the temperature and $\tilde{\tau}_{GR}$ and $\tilde{\tau}_v$
are the fiducial gravitational radiation time scale and the
fiducial viscous time scale respectively. The last two are defined respectively by the  following expressions (for arbitrary value m)
\begin{equation}
\tau_{GR}=\tilde{\tau}_{GR}\left(\frac{\Omega_0}{\Omega}
\right)^{2m+2},  \label{fid-t-GR}
\end{equation}
\begin{equation}
\tau_v=\tilde{\tau}_v\left(\frac{\Omega_0}{\Omega} \right)^{1/2}
\left(\frac{T}{10^8 \ K} \right). \label{fid-t-v}
\end{equation}
%
%. We found that
%\[\Omega_0=3.15487 \cdot 10^5\sqrt{\left(\frac{ M}{M_{\odot}} \right)
%\left(\frac{ {\rm km}}{R} \right)^3}  \]
The gravitational radiation time scale $\tau_{GR}$ is  given by~\cite{Lidblom-2000}
%\begin{widetext}
\begin{eqnarray}
\frac{1}{\tau_{GR}}&=&-\frac{32\pi G
\Omega^{2m+2}}{c^{2m+3}}\frac{(m-1)^{2m}}{[(2m+1)!!]^2}\left(\frac{m+2}{m+1}\right)^{2m+2}
\int_0^{R_{\rm core}}\rho(r) r^{2m+2} dr. \label{tgr-1}
\end{eqnarray}
%\end{widetext}
%
The damping time scale $\tau_{v}$ due to viscous dissipation at the boundary
layer of the perfectly rigid crust and fluid core is given by
\cite{Lidblom-2000}
%\begin{widetext}
\begin{eqnarray}
\tau_{v}&=&\frac{1}{2\Omega}\frac{2^{m+3/2}(m+1)!}{m(2m+1)!! {\cal
I}_m}\sqrt{\frac{2\Omega R_{\rm core}^2\rho_{\rm t}}{\eta_{\rm t}}}
\int_0^{R_{\rm core}}\frac{\rho(r)}{\rho_{\rm t}}\left(\frac{r}{R_{\rm core}}\right)^{2m+2}
\frac{dr}{R_{\rm core}}.
 \label{tv-1}
\end{eqnarray}
%\end{widetext}
%
$\Omega$ is the angular velocity of the unperturbed star,
$\rho(r)$ is the radial dependence of the mass density of the
neutron star, $R_{\rm core}$, $\rho_{\rm t}$ and $\eta_{\rm t}$ are the radius, density
and viscosity of the fluid at the outer edge of the core respectively.
In neutron stars colder than about $10^9$ K the shear viscosity is
expected to be dominated by electron-electron scattering. The
viscosity associated with this process is given by
\cite{Lidblom-2000}
\begin{equation}
\eta_{ee}=6.0\times 10^6 \rho^2 T^{-2}, \qquad ({\rm g \ cm^{-1}}
\ s^{-1}), \label{eta-ee-1}
\end{equation}
where all quantities are given in cgs units and $T$ is measured in
K. For temperature above $10^9$ K, neutron-neutron scattering
provides the dominant dissipation mechanism. In this range the
viscosity is given by \cite{Lidblom-2000}
\begin{equation}
\eta_{nn}=347 \rho^{9/4} T^{-2},\qquad ({\rm g\ cm^{-1}\ s^{-1}}).
\label{eta-nn-1}
\end{equation}
In the present work we consider the case of $m=2$ r-mode and also  we neglect the effects of bulk viscosity,
which are not important for $T \leq 10^{10} \ K  $.
In our  previous work it was found that the time scale $\tilde{\tau}_{GR}$ takes the form~\cite{Moustakidis-015}
\begin{equation}
\tilde{\tau}_{GR}=-0.7429 \left(\frac{R}{ {\rm km}}\right)^9
\left(\frac{ M_{\odot}}{M}\right)^3 \left[I(R_c)\right]^{-1} \ ({\rm s}), \label{taugr-1}
\end{equation}
where
\begin{equation}
I(R_c)=\int_{0}^{R_{\rm core}}\left(\frac{\epsilon
(r)}{{\rm MeV \ fm^{-3} }}\right) \left(\frac{r}{{\rm km}}\right)^6 d\left(\frac{r}{{\rm km}}\right).
\label{Ic-1}
\end{equation}
The integral $I(R_c)$  is  a basic ingredient of the r-mode studies (see Ref.~\cite{Moustakidis-015}).
The fiducial viscous time $\tilde{\tau}_{v}$,  after some algebra, is written for the case of
viscosity due to  electron-electron  and neutron-neutron
scattering respectively~\cite{Moustakidis-015}
\begin{eqnarray}
\tilde{\tau}_{ee}&=&10.8386 \left(\frac{R}{ {\rm
km}}\right)^{3/4}\left(\frac{ M_{\odot}}{M} \right)^{1/4}
\left(\frac{ {\rm km}}{R_{\rm c}} \right)^6
\left(\frac{{\rm MeV \ fm^{-3} }}{{\cal E}_{\rm t}} \right)^{3/2} \ I(R_c) \ ({\rm s}), \label{t-ee-2}
\end{eqnarray}
\begin{eqnarray}
\tilde{\tau}_{nn}&=&41.904
 \left(\frac{R}{
{\rm km}}\right)^{3/4}\left(\frac{ M_{\odot}}{M} \right)^{1/4}
\left(\frac{ {\rm km}}{R_{\rm c}} \right)^6
\left(\frac{{\rm MeV \ fm^{-3} }}{{\cal E}_{\rm t}}\right)^{13/8} \ I(R_c) \ ({\rm s}). \label{t-nn-2}
\end{eqnarray}
The corresponding critical angular velocities  $\Omega_c$ are given by the relation
\begin{eqnarray}
\Omega_c^{ee}&=&1.9377\cdot 10^5\left(\frac{R_c}{{\rm Km}}\right)^{12/11}
\left(\frac{{\cal E}_{\rm t}}{\rm MeV \ fm^{-3} }\right)^{3/11}
\left(I(R_c)\right)^{-4/11}\left(\frac{10^8 \ K}{T}  \right)^{2/11} \ ({\rm s}^{-1})
\label{Omega-c-eos-ee}
\end{eqnarray}
and also
\begin{eqnarray}
\Omega_c^{nn}&=&0.930515\cdot 10^5\left(\frac{R_c}{{\rm Km}}\right)^{12/11}
\left(\frac{{\cal E}_{\rm t}}{\rm MeV \ fm^{-3} }\right)^{13/44}
\left(I(R_c)\right)^{-4/11}\left(\frac{10^8 \ K}{T}  \right)^{2/11} \ ({\rm s}^{-1}).
\label{Omega-c-eos-nn}
\end{eqnarray}
%%%
Now we consider that, in a very good approximation,
the energy density of a neutron star is given by the Tolman VII analytical  solution
\begin{equation}
{\cal E}(r)=\frac{15 M c^2}{8\pi R^3}\left(1-\left(\frac{r}{R}\right)^2\right).
\label{Tolman-1}
\end{equation}
It is well known that despite its simplicity, this  distribution reproduces in a very good accuracy various neutron star properties including
the binding energy and moment of inertia while being  in good agreement with realistic equations of state for neutron stars
with $M >1M_{\odot}$~\cite{Lattimer-01,Ragjoonundun-015}. Moreover, the Tolman VII solution has the correct behavior not only on
the extreme limits $r = 0$ and $r = R$ but also in the intermediate region (see Fig. 5 of Ref.~\cite{Lattimer-01}). Below we will employ the Tolman VII solution, in order to provide some analytical expressions for  the fiducial times and the critical temperature,  for two reasons: a) firstly to exhibit the role played by the crust-core interface and b) to provide some analytical expressions which can be easily manipulated and used  for the study of the r- mode instability windows.
Now, the integral $I(R_c)$ takes the analytical form
\begin{eqnarray}
I(R_c)&=&10583.45\left( \frac{M}{M_{\odot}}\right) \left( \frac{R_{\rm core}}{\rm km}\right)^4\left( \frac{R_{\rm core}}{R}\right)^3
 \left(9-7\left( \frac{R_{\rm core}}{R}\right)^2\right).
\label{Ic-2}
\end{eqnarray}
The above approximation is accurate ($4 \% $ deviation for $M=1.4 M_{\odot}$ and less than $1 \% $ for $M \geq 1.7 M_{\odot}$).
It can be found easily that the use of the Tolman VII solutions leads to
\begin{equation}
M_{\rm core}=\frac{5 M}{2}\left(\frac{R_{\rm core}}{R}\right)^3\left[1-\frac{3}{5}\left(\frac{R_{\rm core}}{R}\right)^2\right].
\label{Mcore-1}
\end{equation}
In addition, we {\bf obtained}   using the approximation~(\ref{Cradii-3}) that
\begin{equation}
M_{\rm core}=\frac{5 M}{2}\left(\frac{2\beta h_{\rm t}}{h_{\rm t}-1+2\beta}\right)^3\left[1-\frac{3}{5}\left(\frac{2\beta h_{\rm t}}{h_{\rm t}-1+2\beta}\right)^2\right].
\label{Mcore-2}
\end{equation}
This approximation is also very accurate ($4 \% $ deviation for $M=1.4 M_{\odot}$ and less than $1 \% $ for $M \geq 1.7 M_{\odot}$).
However, it fails to reproduce with  the proper accuracy the mass of the crust $M_{\rm crust}$.

Since   the fiducial time scales (and also the critical angular momentum)  are functionals of the integral $I(R_c)$ we can proceed to derive analytical solutions,  by replacing  its value using Eq.~(\ref{Ic-2}).
In this case the fiducial time $\tilde{\tau}_{GR}$ takes the form
\begin{eqnarray}
\tilde{\tau}_{GR}&=&-7\cdot 10^{-5} \left( \frac{R}{R_{\rm core}}\right)^7\left( \frac{R}{\rm km}\right)^5 \left(\frac{M_{\odot}}{M}  \right)^4
\left(9-7\left( \frac{R_{\rm core}}{R}\right)^2\right)^{-1}.
\label{taugr-2}
\end{eqnarray}
Taking into account that at  the transition density   it holds $P_{\rm t}\ll {\cal E}_{\rm t}$  and therefore  $\mu_{\rm t}\simeq {\cal E}_{\rm t} /n_{\rm t}$ we find
\begin{equation}
{\cal E}_{\rm t}\simeq \mu_0 n_{\rm t} \sqrt{h_{\rm t}}.
\label{et-1}
\end{equation}
Using the above approximation  the time scales $\tilde{\tau}_{ee}$ and $\tilde{\tau}_{nn}$ are  written respectively
\begin{eqnarray}
\tilde{\tau}_{ee}&=&4.042\left(\frac{{\rm fm^{-3} }}{n_{\rm t}}\right)^{3/2}\frac{1}{h^{3/4}_{\rm t}}
 \left(\frac{R_{\rm core}}{ R}\right)\left(\frac{M }{M_{\odot}} \right)^{3/4}
\left(\frac{ {\rm km}}{R} \right)^{5/4}
\left(9-7\left( \frac{R_{\rm core}}{R}\right)^2\right),
\label{t-ee-3}
\end{eqnarray}
%
%and also
\begin{eqnarray}
\tilde{\tau}_{nn}&=&6.65\left(\frac{{\rm fm^{-3} }}{n_{\rm t}}\right)^{13/8}\frac{1}{h^{13/16}_{\rm t}}
 \left(\frac{R_{\rm core}}{ R}\right)\left(\frac{M }{M_{\odot}} \right)^{3/4}
\left(\frac{ {\rm km}}{R} \right)^{5/4}   \left(9-7\left( \frac{R_{\rm core}}{R}\right)^2\right).
\label{t-nn-3}
\end{eqnarray}
It is worth to present also the following analytical expressions for the critical frequencies
\begin{eqnarray}
\Omega_c^{ee}&=&4.298\times 10^4 \left(\frac{n_{\rm t}}{{\rm fm^{-3} }}\right)^{3/11}
h_{\rm t}^{3/22}
\left(\frac{R_{\rm core}}{R}\right)^{-16/11}
\left(\frac{{\rm km}}{R } \right)^{4/11}
\left(\frac{M_{\odot} }{M} \right)^{4/11}\nonumber \\
&\times&   \left(9-7\left( \frac{R_{\rm core}}{R}\right)^{2}\right)^{-4/11}
\left(\frac{10^8 \ K}{T}  \right)^{2/11} \ ({\rm s}^{-1}),
\label{Omega-c-eos-ee}
\end{eqnarray}
%
%and also
\begin{eqnarray}
\Omega_c^{nn}&=&3.926\times 10^4 \left(\frac{n_{\rm t}}{{\rm fm^{-3} }}\right)^{13/44}
h_{\rm t}^{13/88}
\left(\frac{R_{\rm core}}{R}\right)^{-16/11}
\left(\frac{{\rm km}}{R} \right)^{4/11}
\left(\frac{M_{\odot} }{M} \right)^{4/11}\nonumber \\
&\times&   \left(9-7\left( \frac{R_{\rm core}}{R}\right)^{2}\right)^{-4/11}
\left(\frac{10^8 \ K}{T}  \right)^{2/11} \ ({\rm s}^{-1}).
\label{Omega-c-eos-nn}
\end{eqnarray}
The above  expressions although being approximations,  exhibit the dependence of the instability window on the main properties of the crust-core interface.
Moreover, the maximum angular velocity $\Omega_{\rm K}$ (Kepler angular velocity) for any star occurs when the material at the surface effectively orbits the star~\cite{Lindblom-98}. This velocity is nearly $\Omega_{\rm K}=\frac{2}{3}\Omega_0$. Thus, there is a critical temperature  $T_c$ for which
the gravitational-radiation instability is completely suppressed
by viscosity is given by~\cite{Lidblom-2000}
\begin{equation}
\frac{T_c}{10^8 {\rm K}}=\left(\frac{\Omega_0}{\Omega_c} \right)^{11/2}\left(-\frac{\tilde{\tau}_{GR}}{\tilde{\tau}_v}
\right)=\left(\frac{3}{2}\right)^{11/2}\left(-\frac{\tilde{\tau}_{GR}}{\tilde{\tau}_v}
\right).
\label{Tc-1}
\end{equation}
A decrease of  $T_c$ leads to an increment of the instability window (at least for low values of temperatures).

%%%%%%%%%%%%%%%%%%%%%%%%%%%%%%%%%%%
We discuss briefly  the case of an elastic crust. In this case the r-mode penetrates the crust and consequently
the relative motion (slippage) between the crust
and the core is strongly reduced compared to the rigid crust
limit~\cite{Levin-01,Glampedakis-06,Papazoglou-016}.  In this consideration the slippage factor S has been included
on the r-mode problem and the revised time scale is written
\begin{equation}
\tau_{ee(nn)}^{\cal S}\rightarrow \frac{\tau_{ee(nn)}}{{\cal S}^2}.
\label{Slip-1}
\end{equation}
Actually, the factor ${\cal S}$ depends mainly on the angular velocity $Ù$, the core radius $R_c$ and the shear modulus  but can
be treated also,  approximately, as a constant which is varied in the interval from very low values (${\cal S} = 0.05$) up to  ${\cal S} = 1$
corresponding  to a complete rigid crust.

%%%%%%%%%%%%%%%%%%%%%%%%%%%%%%%%%%%%%
\section{Results and Discussion}
%%%%%%%%%%%%%%%%%%%%%%%%%%%%%%%%%%%
%%%%%%%%%%%%%%%%%%%%%%%%%%%%%%%%%%%%%%%%%%%%%%%%%%%%%%%%%%%%%%%%%%%%%%%%
\subsection{Accuracy of the dynamical approximations and the gradient coefficients  $D_{ij}$ }
%%%%%%%%%%%%%%%%%%%%%%%%%%%%%%%%%%%%%%%%%%%%
Firstly, we check the accuracy of the approximation for the effective interaction $U_{\rm dyn}(n,k)$ given in Eq.~(\ref{Vdyn-appr}) compared with the full expression given in Eq.~(\ref{Vdyn-1}). We employ, as an example,  the MDI-FE model   (with $L=80$ MeV  and  $D=72 \ {\rm MeV  \ fm^5}$)  (actually the results and conclusions are similar for all the employed nuclear models). In particular, we found that using the dynamical potential, given by Eq.~(\ref{Vdyn-1}) that $n_{\rm t}=0.0605057 \ {\rm fm^{-3}} $ and $P_{\rm t}=0.185374 \ {\rm MeV  \ fm^{-3}}$, while using the approximation (\ref{Vdyn-appr}) we found that  $n_{\rm t}=0.0603948 \ {\rm fm^{-3}} $ and $P_{\rm t}=0.184194 \ {\rm MeV  \ fm^{-3}}$. In general we found that, in each case, the error of  the transition density is less  than $0.5 \%$ while for the transition pressure is  less  than $1\%$. We  have also  seen  that,  using the parabolic approximation  for the symmetry energy, the error  of the approximation (\ref{Vdyn-appr}) (compared to the expression~(\ref{Vdyn-1})), concerning the values of $n_{\rm t}$ and $P_{\rm t}$, is also  less  than  $0.5 \%$ and $1\%$ respectively. Finally, we investigated the effect of the gradient term $D_{ij}$ on the crust-core interface. We also note that this result is important since, in  most of the cases, the values of $D_{ij}$ are not included in the nuclear models and must be inserted   by hand.
We found that, for reliable values of $D_{ij}$, the approximations (\ref{Vdyn-appr}) and (\ref{Vdyn-1}) predict  similar results.

%%%%%%%%%%%%%%%%%%%%%%%%%%%%%%%%%%%%%%%%%%%%%%%%%%%%%%%%%%%%%%%%%%%%%%%%
\subsection{Transition densities and transition  pressures for  various models and approximations }
%%%%%%%%%%%%%%%%%%%%%%%%%%%%%%%%%%%%%%%%%%%%%%%%%%%%%%%%%%%%%%%%%%%%%%%%%5
So far, we have considered above, both the approximation~(\ref{Vdyn-appr}) providing a high accuracy, independent of the employed nuclear models (including the nuclear symmetry energy) and the values of the gradient terms.
Now we proceed  with the determination of the transition density and pressure for all the proposed nuclear models. Actually, we use mainly four cases.  First we  use the dynamical method by considering   for the calculation of the proton fraction  Eq.~(\ref{beta-3}) (DYN-FE case hereafter) and  the parabolic approximation Eq.~(\ref{b-equil-2}) (DYN-PA case hereafter).  Second, we employ the thermodynamical method for the calculation of the proton fraction  via  Eq.~(\ref{beta-3}) (THER-FE case hereafter) and  the parabolic approximation Eq.~(\ref{b-equil-2}) (THER-PA case hereafter). It must be noted that  similar effects have been  obtained using  other  nuclear models. It has been found that the predicted results are only qualitatively  different.

In Table~2 and ~3 we present the transition density, pressure and the quantity $h_{\rm t}=\frac{1}{\mu_0^2}\left(\frac{{\cal E}_{\rm t}}{n_{\rm t}}\right)^2 $. In Table~2 we  show our results  by employing both the dynamical and the thermodynamical methods in the framework of the full approximation. The values of  $n_{\rm t}$ calculated by the dynamical method are  lower by $(10-15) \%$ compared with the thermodynamical one. Our results confirm previous calculations~\cite{Xu-09,Hebeler-13,Boquera-17}.  The most distinctive feature is the marked lowering of  the values of the transition pressure $P_{\rm t}^{\rm dyn}$  compared to  $P_{\rm t}^{\rm th}$. As we will see below this has also a pronounced effect on the  neutron star properties which are sensitive  to the values of the critical pressure.

In Table~3 we present results corresponding to the use of the parabolic approximation~(\ref{b-equil-2}). In this case the values of $n_{\rm t}$, $P_{\rm t}$ and $h_{\rm t}$ are higher compared to the use of the full approximation. In particular, the values of $n_{\rm t}$ increase even more by $15-20\%$ compared to the full approximation both in the dynamical and thermodynamical methods. The effects are even more sizeable   concerning the transition pressure $P_{\rm t}$ since its values increase twice or even more compared to the dynamical method. The main conclusion is the following: The use of the dynamical method, in the framework of the full approximation for the symmetry energy,   significantly lowers the values of $n_{\rm t}$ and $P_{\rm t}$ compared to  the thermodynamical method (both in parabolic or full approximation). Now, since many neutron star properties  depend on  the crust-core  interface, one has to carefully take into account the transition point. Below, we  examine both quantitative and qualitative effects of the transition point on a few neutron star properties and evolution processes.

%%%%%%%%%%%%%%%%%%%%%%%%%%%%%%%%%%%%%%%%%%%%%%%%%%%%%%%%%%%%%%%%%%%%%%%%
\subsection{Discussion of the approximation for  $M_{\rm crust}$ and relations with the transition pressure }
%%%%%%%%%%%%%%%%%%%%%%%%%%%%%%%%%%%%%%%%%%%%
Before we proceed with the analysis  of the effects of $P_{\rm t}$ and $n_{\rm t}$ on various neutron star static and dynamical properties it is important  to  discuss further the approximations concerning the crustal radius and crustal mass. The approximations (\ref{Cradii-2}) and  (\ref{Cradii-3}) are very accurate and will be used below to derive some analytical expressions for the thermal relaxation time, the QPOs frequencies and the critical angular velocities.

In the present work we derived also a semi-theoretical expression which relates the transition pressure $P_{\rm t}$ with the total radius of a neutron star with mass  $M=1.4 \ M_{\odot}$. Actually, the expression (\ref{Pt-anal-1}) works with a proper accuracy excluding only the very stiff and very soft EoSs.   Since the majority of the observational neutron stars has a mass close to this limit,  the expression (\ref{Pt-anal-1}) may be proven useful in order to construct a {\it bridge} between the bulk observation quantity  $R$ and the microscopic one $P_{\rm t}$. In particular, the accurate observational measurement  of $R$ may help to constraint $P_{\rm t}$. For example a measurement  $R_{1.4}=13$ km will provide the constraint $P_{\rm t}=0.4\pm 0.11 \ {\rm MeV} \ {\rm fm}^{-3}$. Moreover, an accurate experimental measurement of $P_{\rm t}$  may help to constrain the radius. For example the value  $P_{\rm t}=0.45 \ {\rm MeV} \ {\rm fm}^{-3}$  will provide the constraint $R=12.51\pm 0.86$ km.

Now we discuss further the semi-theoretical expression~(\ref{Pt-L}). According to Tables 2 and 3 the use of  the dynamical (thermodynamical) method in the framework of the full approximation satisfies  the expression~(\ref{Pt-L}). However, the use of the parabolic approximation leads to the inverse behavior (see also Ref.~\cite{Xu-09}). This is an additional indication that the PA may  lead to misleading results concerning the values of the transition pressure and its dependence of the slope parameter $L$.

%%%%%%%%%%%%%%%%%%%%%%%%%%%%%%%%%%%%%%%%%%%%%%%%%%%%%%%%%%%%%%%%%%%%%%%%
\subsection{Effects on the frequencies of QPOs and thermal relaxation of the crust   }
%%%%%%%%%%%%%%%%%%%%%%%%%%%%%%%%%%%%%%%%%%%%%%%%%%%%%%%%%%%

In Fig.~2 we present a mass-radius diagram showing the constraints from neutron star seismology (originating from the soft gamma-ray repeater SGR 1806-20)
(for more details see Ref.~\cite{Lattimer-07,Samuelsson-07}). First, we plot the mass-radius dependence using the MDI model (for $L=80$ MeV). We consider that $f_{n=0,l=2}=29$ Hz and also $f_{n=1,l=1}=626.5$ Hz~\cite{Lattimer-07} and we solve Eqs~(\ref{freq-1}) and (\ref{freq-2}) correspondingly for the four selected cases. The predicted mass-radius constraints have been included also in Fig.~2.

Obviously the effects of the transition density are more pronounced in the case of the  $f_{n=1,l=1}$ modes. In particular, the use of the DYN-FE  decreases the corresponding values of the mass (for fixed values of the radius). Moreover, even for the same approximation (FE or PA)  the dynamical method decreases also the constrained values of the mass. As a general conclusion  a more realistic EoS (DYN-FE) decreases appreciably the values of the mass.  In the case of the fundamental mode $f_{n=0,l=2}$, the effects of the EoS are less important and appear mainly for high values of the radius. In the same figure we plot the four values of $\beta$ which emerge from the elimination of $R$  in Eqs.~(\ref{freq-1}) and (\ref{freq-2}). Once again, we found that  the effects of the crust-core interface must be taken into account in order to impose constraints on the mass-radius diagram.

In Fig.~3(a) we present  the  thermal relaxation time  of the crust $t_{\rm w}$  as a function of the neutron star mass $M$ using Eq.~(\ref{Trel-1}) with $t_1=28\pm 0.2 \ {\rm yrs}$ by displaying the results for the four selected cases. The use of the dynamical method with the full approximation leads to  a marked  lowering of the values of  $t_{\rm w}$. As expected the effects are more pronounced for low neutron star mass  due to the strong $t_{\rm w}-R_{\rm crust}$ dependence. It is very interesting to see that the thermodynamical method with  the parabolic approximation  leads to  very high values of $t_{\rm w}$ (more than twice for low masses) compared to the DYN-FE case and consequently to a large error. In Fig.~3(b) we display the constraints of the thermal relaxation time on the R-M diagram for the four selected cases. We consider that $t_1=28\ {\rm yrs}$ and for the three values $t_{\rm w}=3, 10, 30 \ {\rm yrs}$ we solve Eq.~(\ref{Trel-1}) in order to display the M-R dependence for the four selected cases. Obviously the constraints on the R-M diagram imposed by the crust-core interface are important. In particular the use of the realistic DYN-FE method leads to smaller  values for  the neutron star mass $M$ and consequently larger values for the radius $R$ especially in the case of high values of the relaxation time. For low values of $t_{\rm w}$ the effects are less important but not negligible.

%%%%%%%%%%%%%%%%%%%%%%%%%%%%%%%%%%%%%%%%%%%%%%%%%%%%%%%%%%%%%%%%%%%%%%%%
\subsection{The effects on the crustal moment of inertia   }
%%%%%%%%%%%%%%%%%%%%%%%%%%%%%%%%%%%%%%%%%%%%%%%%%%%%%%%%%%%%%
The effects of the transition pressure $P_{\rm t}$  and density $n_{\rm t}$ are important also for the calculations of the crustal moment of inertia. Actually the major dependence is upon the pressure $P_{\rm t}$. According to Eq.~(\ref{inertia-1}) smaller values of $P_{\rm t}$ reduce the crustal moment of inertia leading to more restrictive constraints~\cite{Link-99}. In Fig.~4(a) we display  the fraction $I_{\rm crust}/I$ as a function of the total mass for the MDI model (for $L=80$ MeV),  for the four considered cases.  Obviously, the use of the DYN-FE model (which leads to low values of $P_{\rm t}$) decreases the allowed region compared to the other three cases. In order to clarify further this point in Fig.~4(b) we plot also the constraint $I_{\rm crust}/I\geq 0.14$ for the four cases. In the same figure we display also the M-R dependence for a few selected nuclear models. In any case, the constraints  imposed by the DYN-FE model are the most restrictive and  in any case support the statement that the transition density and pressure must be calculated with the proper accuracy in order to impose reliable constraints on the bulk neutron star properties. The same conclusions will be inferred if one uses even higher values of the ratio  $I_{\rm crust}/I$.

More recently the authors in Ref.~\cite{Andersson-012,Chamel-012} considered that due to entrainment of superfluid neutrons in the crust, the lower limit of $I_{\rm crust}/I$ must be larger, that is $I_{\rm crust}/I>0.07$, in order to explain glitches. From another point of view, Link~\cite{Link-014} discussed in more detail the origin and the connection of the  moment of inertia of the crust and the core  concluding that  low values of $I_{\rm crust}/I$ must be expected. In any case, further observation measurements  of glitches and more refined theoretical calculations   will impose more accurate limits and help to restrict also the crust-core properties.

%%%%%%%%%%%%%%%%%%%%%%%%%%%%%%%%%%%%%%%%%%%%%%%%%%%%%%%%%%%%%%%%%%%%%%%%
\subsection{The effects on the r-mode instabilities    }
%%%%%%%%%%%%%%%%%%%%%%%%%%%%%%%%%%%%%%%%%%%%
In Tables~4 and~5 we present the fiducial time scales as well as the corresponding critical frequencies and the critical temperatures for neutron stars with  $M = 1.4 M_{\odot}$ and  $M = 1.8 M_{\odot}$ respectively. In the same table we include also the results of the approximation due to the use of the Tolman VII analytical solution. The fiducial time scales, especially the viscous time $\tilde{\tau}_{v}$ is sensitive to the employed approximation. Specifically,  the DYN-FE decreases the absolute value of  $\tilde{t}_{GR}$ around $10\%$ and increases  the value of $\tilde{\tau}_{v}$  around twice compared to  the THER-PA (for a neutron star mass $M=1.4 M_{\odot}$).

The effects on the fiducial time scales are well reflected in the values of the critical frequencies $f_c$
as exhibited in Tables 4 and 5. There is also a decrease of the values of $f_c$ between $12-15\%$. This difference is important since as we shall see below there are  some cases of neutron stars which lie close to the limit of the proposed  instability window. In the same Tables we present also the critical temperature $T_c$. The use of the DYN-FE also reduces to double  the values of $T_c$ and consequently increases the instability window at least at low temperatures. It is also noted that  the use of the Tolman VII solution  leads to a  good accuracy of the mentioned quantities (time scales, critical frequencies and temperatures). Actually in our previous work the uniform density approximation has been employed~\cite{Moustakidis-015}. The present results  indicate that, at least in this kind of calculations, the Tolman VII solution  produces  results which  are  in better agreement with those of  realistic calculations compared to the use of the uniform density approximation.

In any case  the DYN-FE leads to a decrease of the critical frequency. To clarify further this point
in Figs.~5(a) and 5(b) we compare the r-mode instability window for the selected four cases with those of the observed neutron stars in low-mass x-ray
binary  (LMXB) and millisecond radio pulsars (MSRPs) for $M = 1.4 M_{\odot}$ and  $M = 1.8 M_{\odot}$ respectively. We find that the instability window drops by $20-40\%$ Hz when the mass is raised from  $M = 1.4 M_{\odot}$ to $M = 1.8 M_{\odot}$. Furthermore, the stiffness of the EoS leads to an increase of the instability window
(which is specified, in this case, by the $f_c-T$ dependence). Following the study of Wen et al.~\cite{Wen-12} and Haskell et al.~\cite{Haskel-12} we include many cases of LMXBs and a few of MSRPs (for more details, see~\cite{Watts-08,Keek-10} and Table 1 of Ref.~\cite{Haskel-12}). The masses of the mentioned stars are not measured accurately. In addition, we point out that  the estimates of the core temperature $T$ have large uncertainties. In the present work, the  values of T are taken from Ref.~\cite{Haskel-12} and the uncertainties, in a few relevant cases, are derived by employing the method suggested in Ref.~\cite{Ho-11}.

It is obvious from Figs.~5(a) and 5(b) that the majority of the stars lie outside the instability windows predicted by the present models. There are four exceptions, that is, the 4U 1608-52, the SAX J1750.8-2900, the 4U-1636-536, and the MXB 1658-298 which  lie close to the instability window (for mass $M=1.4 \ M_{\odot}$) and two of them inside (for mass $M=1.8 \ M_{\odot}$).  In any case, the  stiffness of the EoS has a strong effect on the width of the instability window and this effect is more pronounced for high values of the neutron star mass.

%%%%%%%%%%%%%%%%%%%%%%%%%%%%%%%%%%%%%%%%%%%%%%%%%%%%%%%%%%%%%%%%%%%%%%%%%%5
\section{Concluding remarks }
%%%%%%%%%%%%%%%%%%%%%%%%%%%%%%%%%%%%%%%%%%%%%%%%%%%%%%%%%%%%%%%%%%%%%%%%%%%%%%%
The values of the density, pressure and energy density of the crust-core interface (which strongly depends on the applied  equation of state) play important role on some static and dynamical properties and processes  of neutron stars. The transition pressure is directly related to the crustal mass while the radius of a neutron star can be determined, with a moderate accuracy, by the knowledge of $P_{\rm t}$. Having precise values of $n_{\rm t}$ and ${\cal E}_{\rm t}$, for a neutron star with fixed mass and radius,
the core radius can be determined with high accuracy.   In the present work a semi-analytical expression, based on theoretical and empirical arguments,  has been derived  and presented for $P_{\rm t}$. To be more specific,  a model-independent correlation between $P_{\rm t}$ and the slope parameter $L$ has been  obtained, for  fixed values of the symmetry energy at the saturation density.
The value of the thermal relaxation time of the crust during the cooling process  as well  as  the frequencies  of the crust are  sensitive also  to the crustal thickness and total radius of the star and consequently to the crust-core interface. We found that  even for  the same model the value of $t_{w}$ is significantly reduced, especially for a low mass neutron star if one employs  the DYN-FE method (compared with the THER-PA). Moreover, the use of the DYN-FE leads to an appreciable  decrease of the predicted mass (for given values of radius) compared to the THER-PA method for the $f_{n=1,l-1}$ frequency. The use of the DYN-FE shrinks   the allowed region in a M-R diagram due to corresponding lower values of the crustal moment of inertia. Finally, there is a moderate dependence  of the  critical frequency on  $n_{\rm t}$ and $P_{\rm t}$.
In particular, the use of the DYN-FE method  enlarges the instability window and consequently increases the possibility that  neutron stars are sources of gravitational waves, via the r-mode instability.  Moreover, we employ the Tolman VII analytical solution of the TOV equations to find analytical expressions for the critical frequencies and the relative time scales, for the r-mode instability, in comparison with the numerical predictions. The dependence  of the above quantities on the transition density and energy density   has been presented and compared with the corresponding numerical studies.
The above conclusions strongly  indicate   that the observational determination of the crustal thickness, crustal moment of inertia, thermal relaxation time, QPOs frequencies and critical frequencies would help significantly to constraint the EoS of neutron star close to the crust-core interface and vise-versa.

A final comment is appropriate. The main  motivation of  the present work is not only  to  study in detail possible constraints on the EoS from the crust-core interface since many studies have been dedicated  to this effort. Actually we intended also  to exhibit the extent of sensitivity of the EoS  constraints to the crust-core interface properties (density, pressure, chemical potential e.t.c). We focused on the effects of the error  introduced by employing the parabolic approximation in the framework of  the dynamical and thermodynamical approximation. We estimated that although the PA is an accurate approximation for the total energy per baryon of nuclear matter,  its derivative (which is involved in the calculations of $n_{\rm t}$ and $P_{\rm t}$ via the symmetry energy)  is not. Consequently the deviations from the use of FE are   important and must be taken into account. In total, our findings support  the statement that the  location of the crust-core interface  must be estimated with a high  accuracy so that the imposed constraints on the EoS can be as much as possible  reliable.
%%%%%%%%%%%%%%%%%%%%%%%%%%%%%
\section*{Acknowledgments}
%%%%%%%%%%%%%%%%%%%%%%%%%%%%%%
One of the authors (Ch.C.M)  would like to thank the Theoretical
Astrophysics Department of the University of Tuebingen, where part
of this work was performed  and Professor K. Kokkotas  for his
useful comments on the preparation of the manuscript. The authors would like to thank Dr. Kai Hebeler for providing   the parametrization of the HLPS model
and  Professor C.P. Panos for useful comments on the manuscript.
This work was partially supported by the  COST action PHAROS (CA16214) and the DAAD Germany-Greece grant ID 57340132.

%%%%%%%%%%%%%%%%%%%%%5
\section{Appendix}
%%%%%%%%%%%%%%%%%%%%%
Considering that $E\equiv E_b(n,x)$ is the energy per particle of nuclear matter then the chemical potentials of neutrons and protons are given by the relations
\begin{equation}
\mu_n=E_b+n\left(\frac{\partial E_b}{\partial n}\right)-x\left(\frac{\partial E_b}{\partial x}\right),
\label{Ap-mun-1}
\end{equation}
 \begin{equation}
\mu_p=E_b+n\left(\frac{\partial E_b}{\partial n}\right)+(1-x)\left(\frac{\partial E_b}{\partial x}\right).
\label{Ap-mup-1}
\end{equation}
Also we have
\begin{equation}
\frac{\partial \mu_p}{\partial n_p}=\frac{\partial \mu_p}{\partial n}+\frac{1-x}{n}\frac{\partial \mu_p}{\partial x},
\label{ap-dmup}
\end{equation}
\begin{equation}
\frac{\partial \mu_n}{\partial n_n}=\frac{\partial \mu_n}{\partial n}-\frac{x}{n}\frac{\partial \mu_n}{\partial x},
\label{ap-dmun}
\end{equation}
\begin{equation}
\frac{\partial \mu_p}{\partial n_n}=\frac{\partial \mu_p}{\partial n}-\frac{x}{n}\frac{\partial \mu_p}{\partial x}.
\label{ap-dmunp}
\end{equation}
After some algebra we find
%\begin{widetext}
\begin{eqnarray}
\frac{\partial \mu_n}{\partial n_n}&=&2\left(\frac{\partial E_b}{\partial n}\right)+n\left(\frac{\partial^2 E_b}{\partial n^2}\right)
-2x\left(\frac{\partial^2 E_b}{\partial n \partial x}\right)+\frac{x^2}{n}\left(\frac{\partial^2 E_b}{\partial x^2}\right),
\label{ap-dmunnn}
\end{eqnarray}
\begin{eqnarray}
\frac{\partial \mu_p}{\partial n_p}&=&2\left(\frac{\partial E_b}{\partial n}\right)+n\left(\frac{\partial^2 E_b}{\partial n^2}\right)
+2(1-x)\left(\frac{\partial^2 E_b}{\partial n \partial x}\right)+\frac{(1-x)^2}{n}\left(\frac{\partial^2 E_b}{\partial x^2}\right),
\label{ap-dmppp}
\end{eqnarray}
\begin{eqnarray}
\frac{\partial \mu_p}{\partial n_n}&=&2\left(\frac{\partial E_b}{\partial n}\right)+n\left(\frac{\partial^2 E_b}{\partial n^2}\right)
+(1-2x)\left(\frac{\partial^2 E_b}{\partial n \partial x}\right)-\frac{x(1-x)}{n}\left(\frac{\partial^2 E_b}{\partial x^2}\right).
\label{ap-dmpnn}
\end{eqnarray}

%%%%%%%%%%%%%%%%%%%%%%%%%%%%%%%%%%%%%%%%%%%%%%%%%%%%%%%%%%%%%%%%%%%%

%\end{widetext}

%%%%%%%%%%%%%%%%%%%%%%%%%%%%%%%

%%%%%%%%%%%%%%%%%%%%%%%%%%%%%%%%%%%%%%%%%%%%%%%%%%%%%%%%%%%%%%%%%%%%%%
%FIGURE-5
%\begin{figure}
%\centering
%\includegraphics[height=8.1cm,width=8.1cm]{Graph3.eps}\
%\includegraphics[height=6.1cm,width=7.1cm]{fig-2b.eps}\
%\caption{The mass-radius relations for the selected EOSs }
%\label{}
%\end{figure}
%%%%%%%%%%%%%%%%%%%%%%%%%%%%%%%%%%%%%%%%%%%%%%%%%%%%%%%%%%%%%%%%%%%%%%
%%%%%%%%%%%%%%%%%%%%%%%%%%%%%%%%%%%%%%%%%%%%%%%%%%%%%%%%%%%%%%%%%%%%%%
%FIGURE-5
\begin{figure}
\centering
\includegraphics[height=8.cm,width=9.5cm]{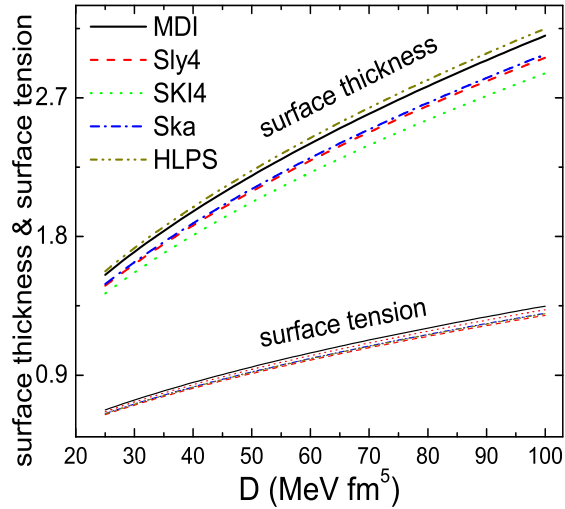}\
\caption{The thickness parameter $t_{90-10}$ (in fm) and the surface tension $\sigma_{\rm snm}$ (in MeV/fm$^{2}$)  as a function of the parameter  $D$ for  the employed models. }
\label{}
\end{figure}
%%%%%%%%%%%%%%%%%%%%%%%%%%%%%%%%%%%%%%%%%%%%%%%%%%%%%%%%%%%%%%%%%%%%%%

%%%%%%%%%%%%%%%%%%%%%%%%%%%%%%%%%%%%%%%%%%%%%%%%%%%%%%%%%%%%%%%%%%%%%%
%FIGURE-5
\begin{figure}
\centering
\includegraphics[height=8.5cm,width=8.5cm]{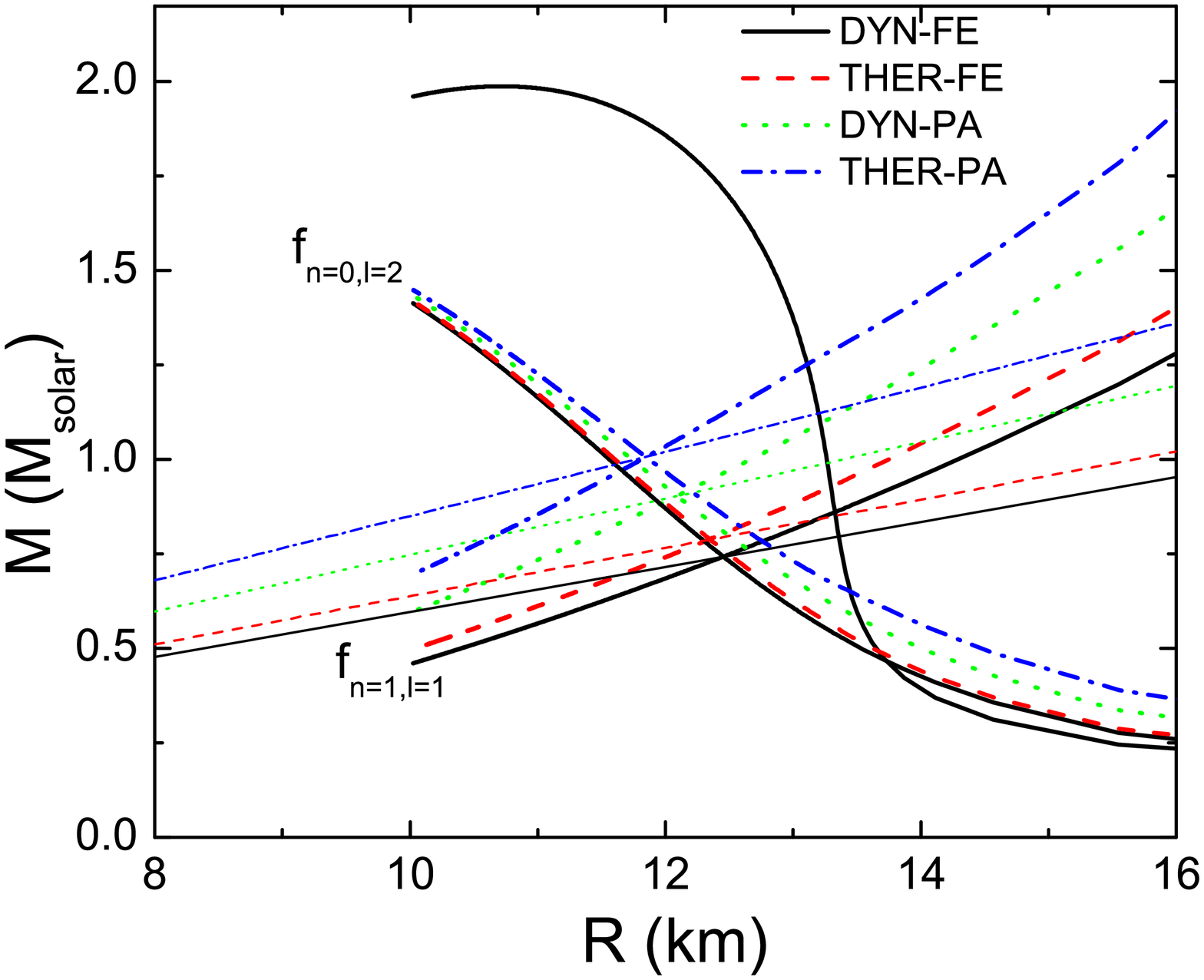}\
\caption{The mass as a function of the radii for the MDI model (for $L=80$ MeV) including constraints from neutron star seismology according to Eqs.~(\ref{freq-1}) and (\ref{freq-2}) and for the four selected cases. The straight lines  correspond to the equation $\beta=\beta(h_{\rm t})$, which emerges  from the elimination of $R$ in Eqs.~(\ref{freq-1}) and (\ref{freq-2}), for each of the four cases. }
\label{}
\end{figure}
%%%%%%%%%%%%%%%%%%%%%%%%%%%%%%%%%%%%%%%%%%%%%%%%%%%%%%%%%%%%%%%%%%%%%%

%%%%%%%%%%%%%%%%%%%%%%%%%%%%%%%%%%%%%%%%%%%%%%%%%%%%%%%%%%%%%%%%%%%%%%
%FIGURE-5
\begin{figure}
\centering
\includegraphics[height=8.1cm,width=8.1cm]{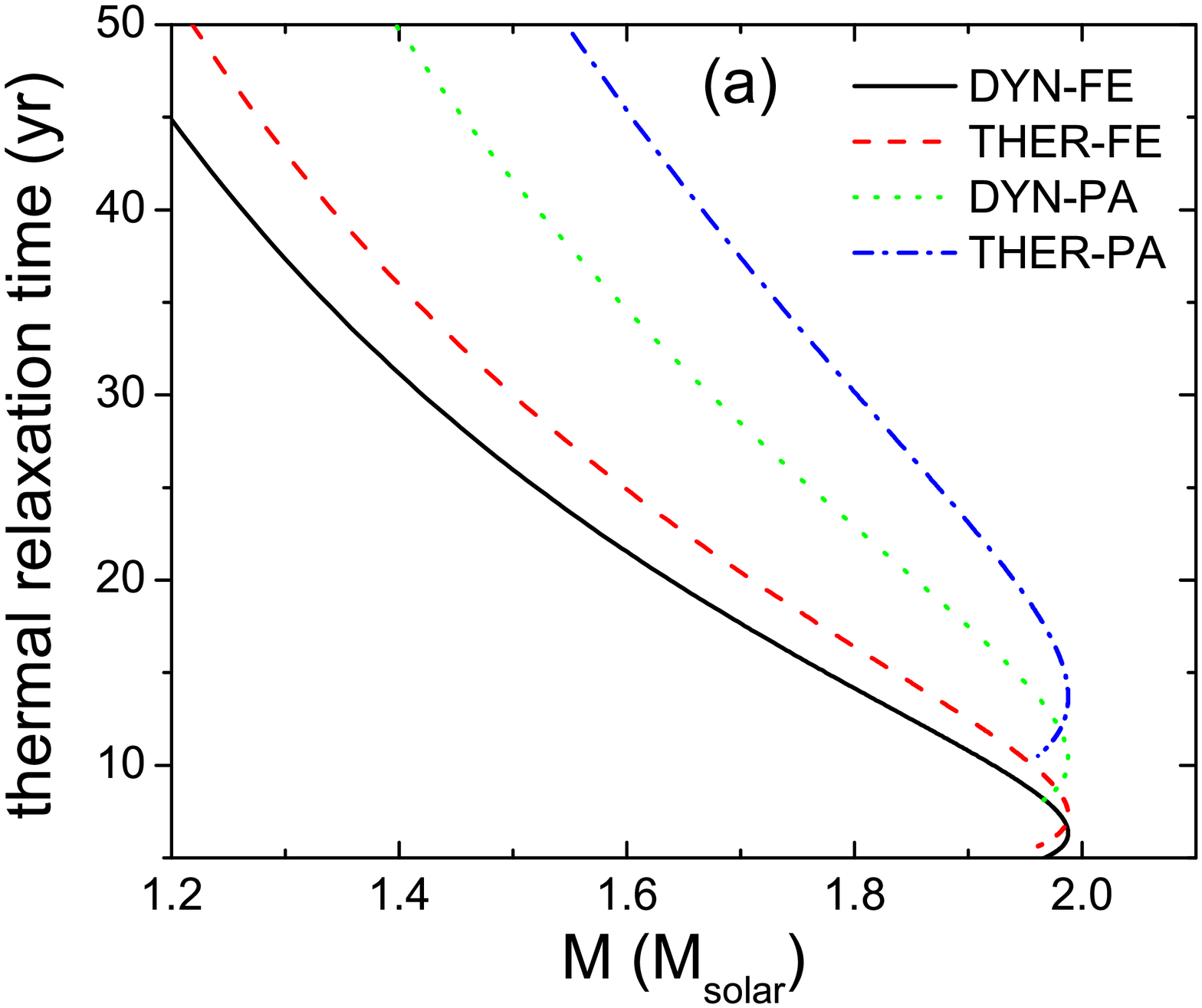}\
\includegraphics[height=8.1cm,width=8.9cm]{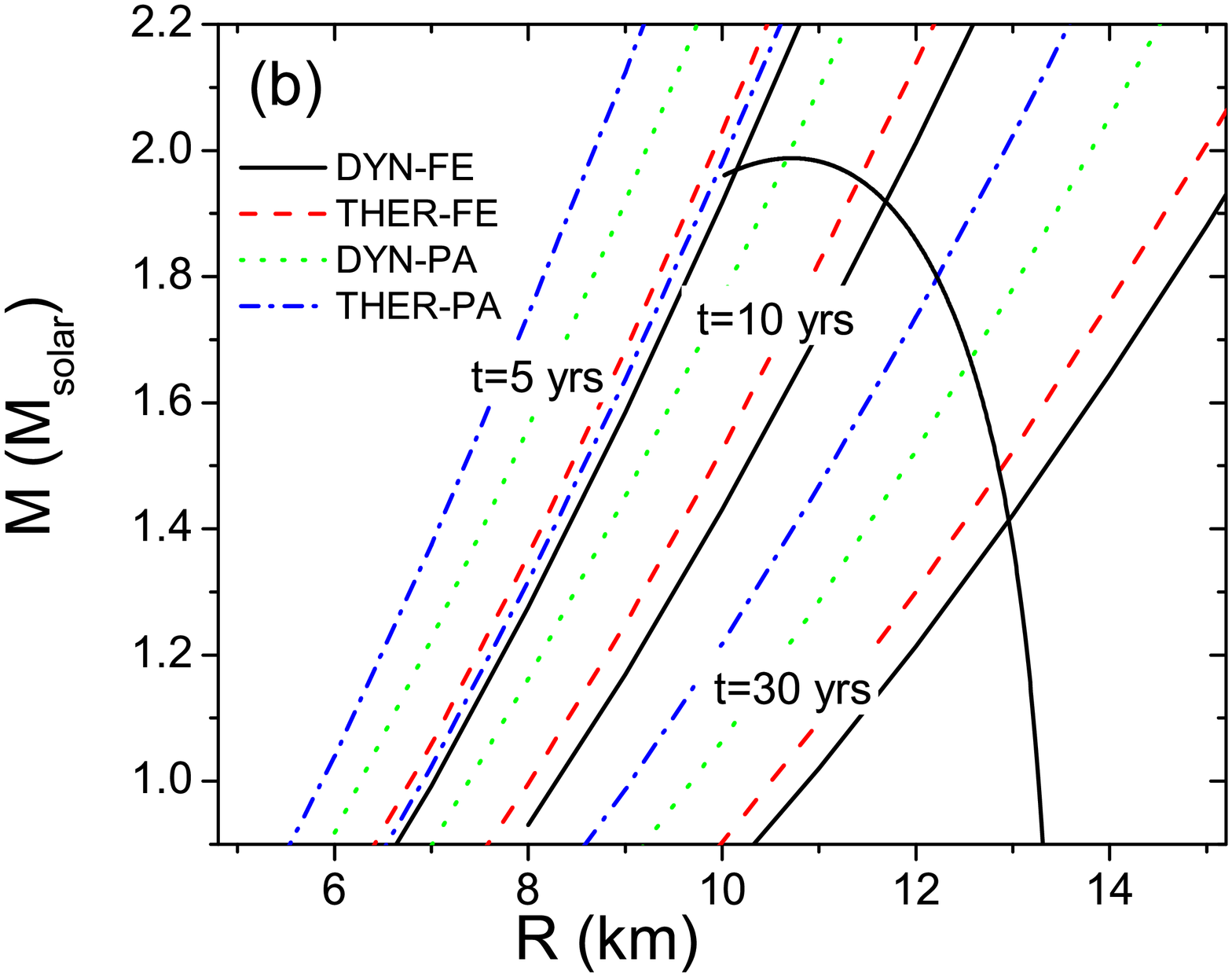}\
\caption{(a) The thermal relaxation time $t_w$ as a function of the total mass for the four selected cases. (b) Constraints on the M-R diagram form the thermal relaxation time $t_w$ for the four selected cases. The M-R dependence for the MDI model ($L=80$ MeV) has been included also for comparison. For more details see text.    }
\label{}
\end{figure}
%%%%%%%%%%%%%%%%%%%%%%%%%%%%%%%%%%%%%%%%%%%%%%%%%%%%%%%%%%%%%%%%%%%%%%
%%%%%%%%%%%%%%%%%%%%%%%%%%%%%%%%%%%%%%%%%%%%%%%%%%%%%%%%%%%%%%%%%%%%%%
%FIGURE-5
\begin{figure}
\centering
\includegraphics[height=8.1cm,width=8.5cm]{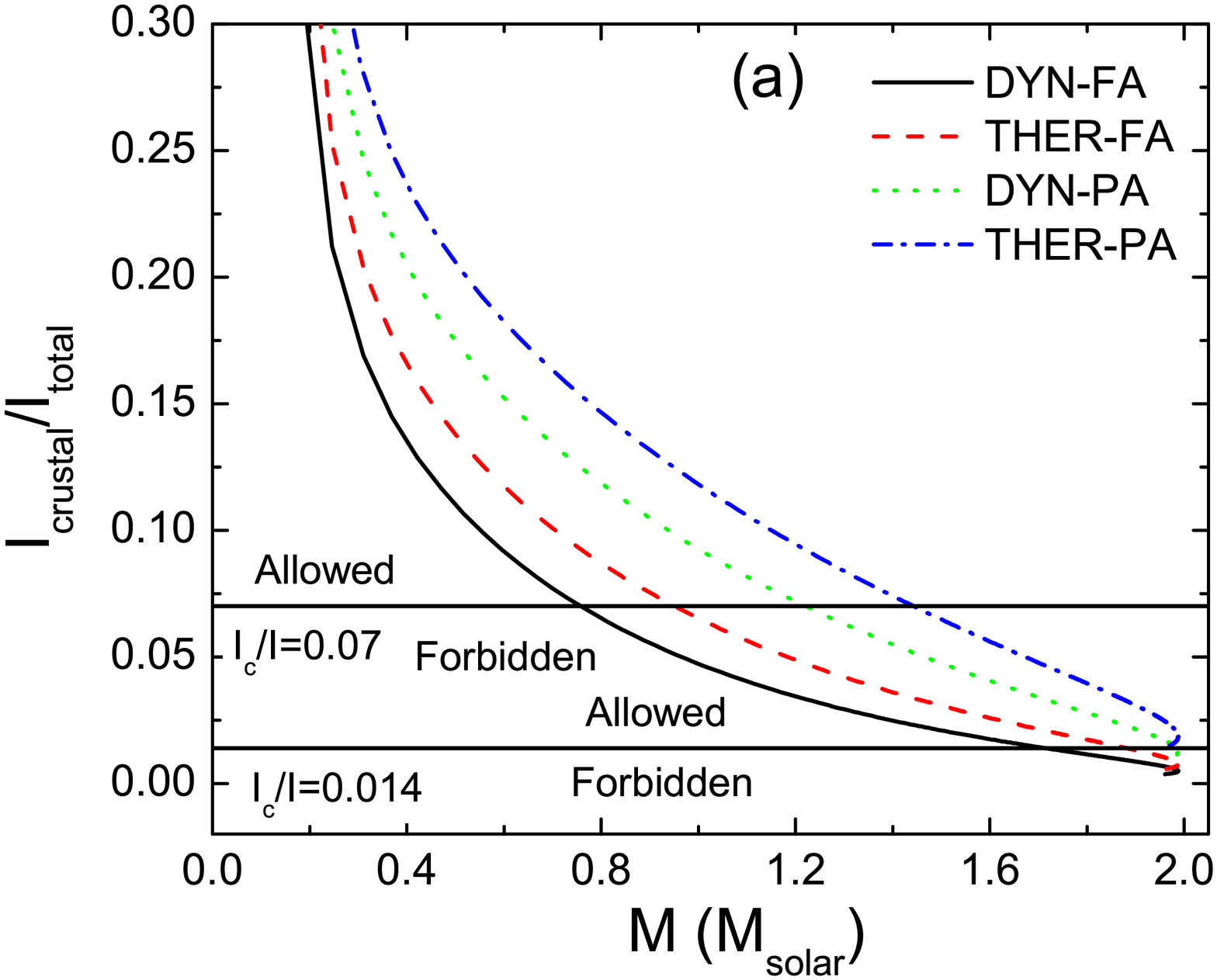}\
\includegraphics[height=8.1cm,width=8.5cm]{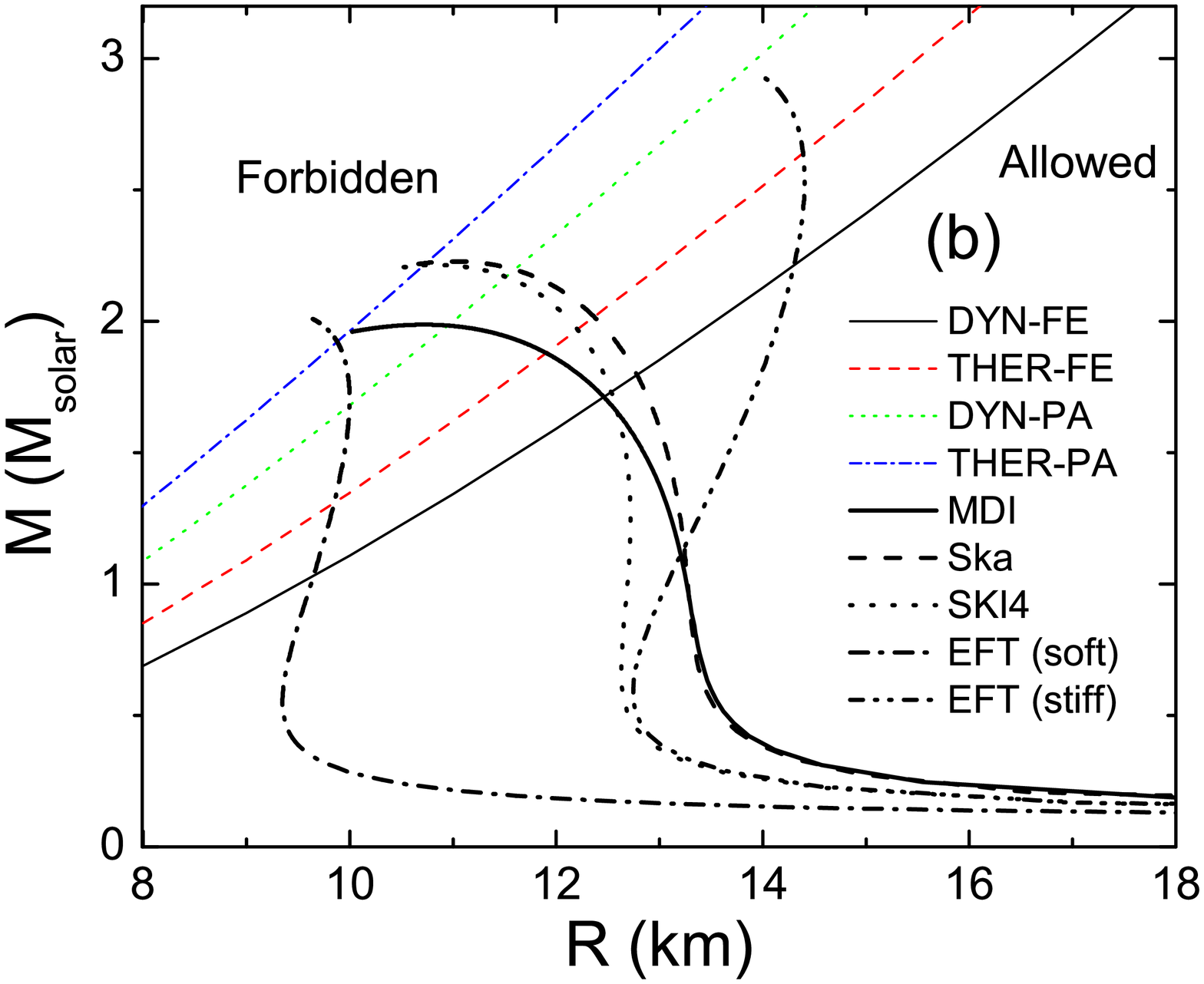}\
\caption{ (a) The crustal fraction of the moment of the inertia as a function of mass presented  for the four selected cases. For comparison
we include the horizontal lines, each one representing a possible $I_{\rm crust}/I$  constraint, deduced for the Vela pulsar (assuming a mass  $M=1.4 M_{\odot}$).
(b) The mass-radius diagram for various nuclear EoS and the constraints $I_{\rm crust}/I=0.014$ are derived from  the four selected cases. }
\label{}
\end{figure}
%%%%%%%%%%%%%%%%%%%%%%%%%%%%%%%%%%%%%%%%%%%%%%%%%%%%%%%%%%%%%%%%%%%%%%
%%%%%%%%%%%%%%%%%%%%%%%%%%%%%%%%%%%%%%%%%%%%%%%%%%%%%%%%%%%%%%%%%%%%%%
%FIGURE-5
\begin{figure}
\centering
\includegraphics[height=8.1cm,width=8.5cm]{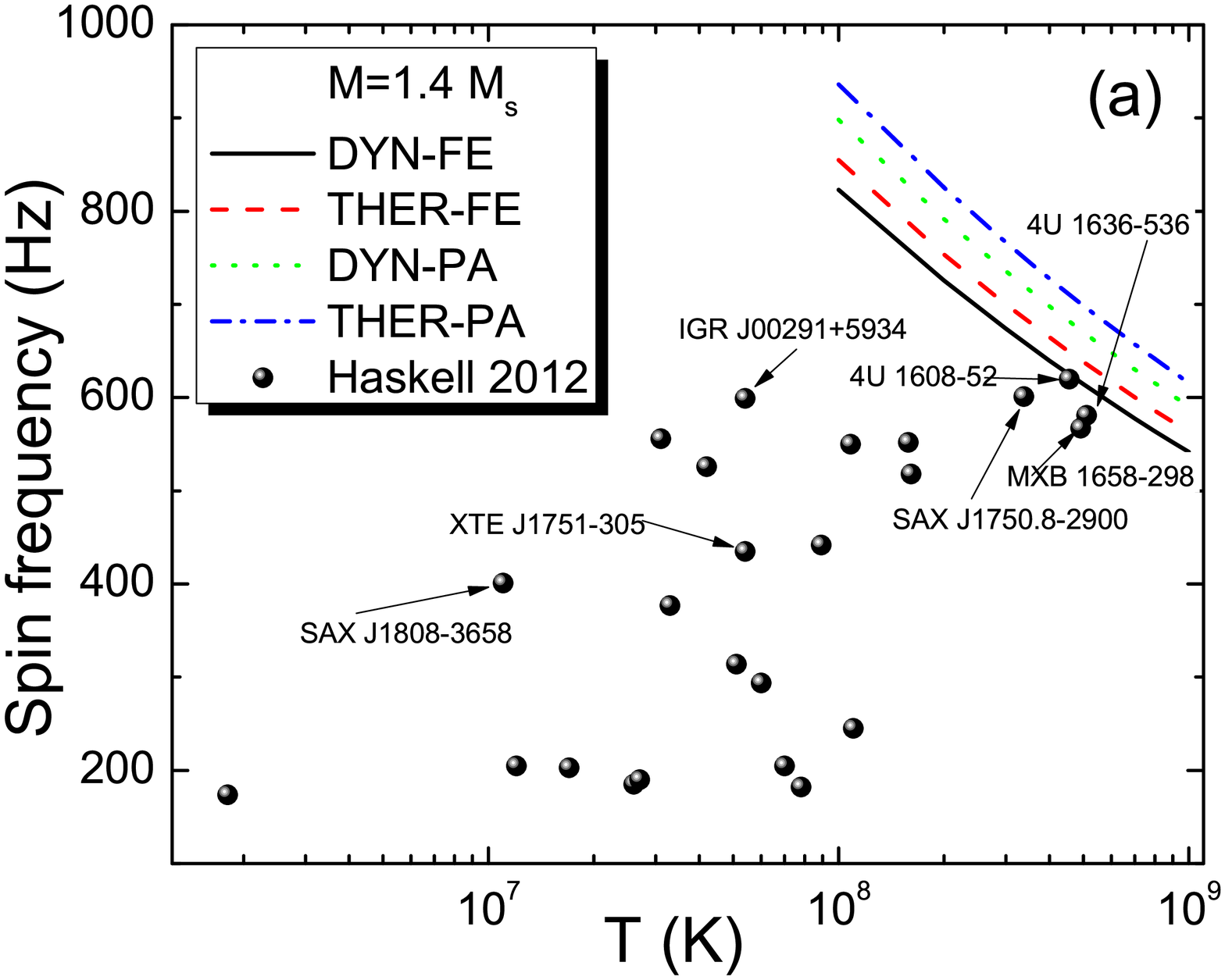}\
\includegraphics[height=8.1cm,width=8.5cm]{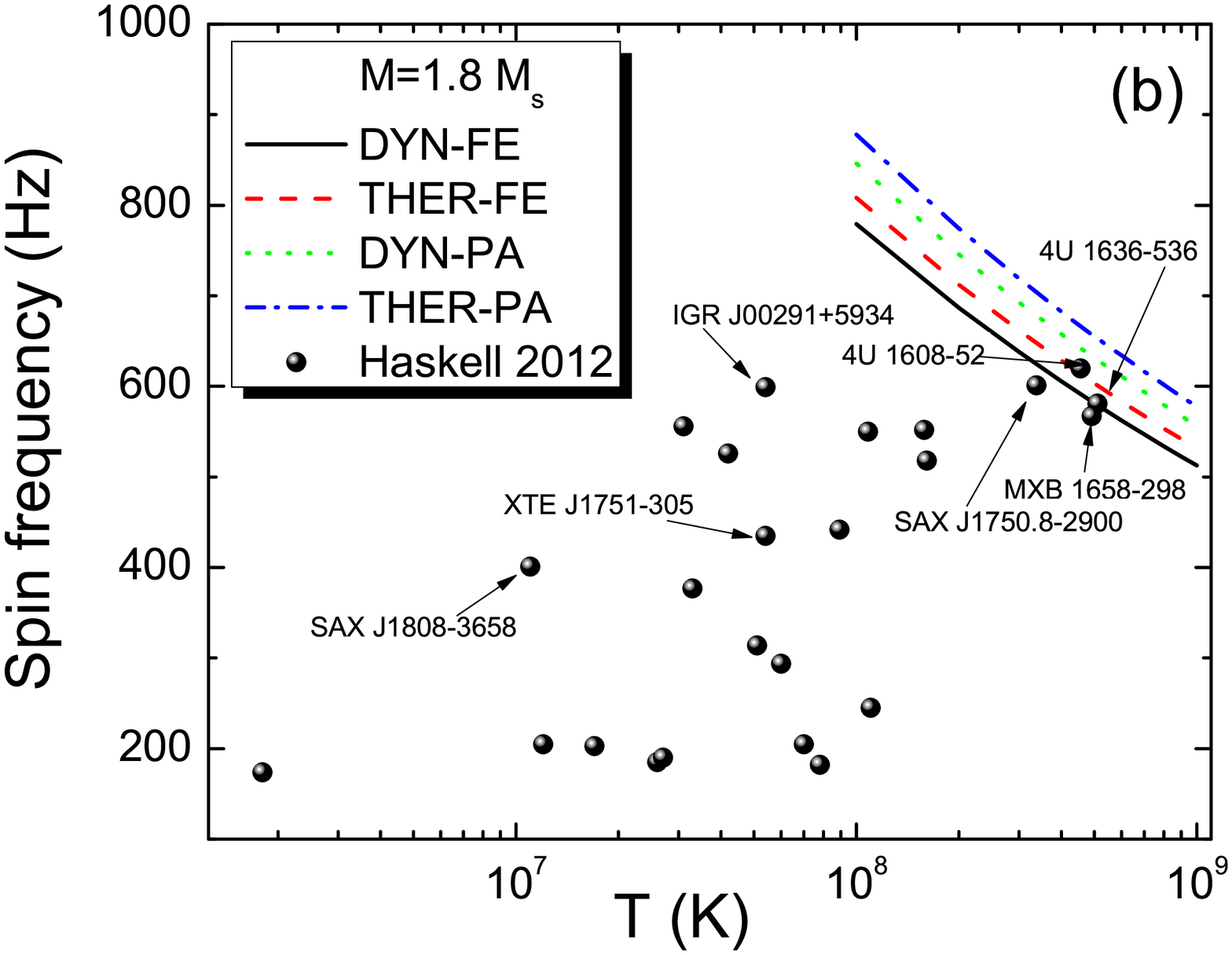}\
\caption{The critical frequency-temperature dependence for a neutron star with mass  $M = 1.4 M_{\odot}$ (a) and $M = 1.8 M_{\odot}$ (b)
constructed for the selected EOSs. The observed cases of LMXBs and MSRPs from Haskell {\it et al}.~\cite{Haskel-12} are also included for a comparison.  The cases IGR J00291+5934, XTE J1751-305, and
SAX J1808-3658 with well-known observation spin-down rate, are also indicated. }
\label{}
\end{figure}
%%%%%%%%%%%%%%%%%%%%%%%%%%%%%%%%%%%%%%%%%%%%%%%%%%%%%%%%%%%%%%%%%%%%%%

%%%%%%%%%%%%%%%%%%%%%%%%%%%%%%%%%%%%%%%%%%%%%%%%%%%%%%%%%%%%%%%%%%%%%%
\begin{table}[h]
 \begin{center}
\caption{ The values of the thickness parameter $t_{90-10}$ (in fm) and the surface tension $\sigma_{\rm snm}$ (in MeV/fm$^{2}$)  for $D=72$ MeV fm$^{5}$ derived for the employed models.}
\vspace{0.5cm}
\begin{tabular}{|l|c|c|c|c|c|c|c|c|c|c|c|}
\hline
                  & MDI &  Sly4      &     SKI4  &Ska   & HLPS    \\
\hline
$t_{90-10}$       &   2.63       &   2.51    &  2.43     & 2.52     & 2.67        \\
\hline
 $\sigma_{\rm snm}$   &   1.14      &   1.09    &  1.12     & 1.10     &  1.10     \\
\hline
\end{tabular}
\end{center}
\label{t-sigma}
\end{table}
%%%%%%%%%%%%%%%%%%%%%%%%%%%%%%%%%%%%%%%%%%%%%%%%%%%%%%%%%%%%%%%%%%%%%%%%%%%%%%%

%%%%%%%%%%%%%%%%%%%%%%%%%%%%%%%%%%%%%%%%%%%%%%%%%%%%%%%%%%%%%%%%%%%%%%

\begin{table}[h]
 \begin{center}
\caption{The transition density $n_{\rm t}$ (in fm$^{-3}$),  pressure
$P_{\rm t}$ (in MeV/fm$^3$) and the thermodynamical factor $h_{\rm t}$ obtained from the considered models  by employing
the full expansion. } \label{t:2}
\vspace{0.5cm}
\begin{tabular}{|l|c|c|c|c|c|c|}
\hline
Model &  $n_{\rm t}^{\rm dyn}$  & $P_{{\rm t}}^{{\rm dyn}} $     & $h_{\rm t}^{\rm dyn}$   & $n_{{\rm t}}^{{\rm th}}$     & $P_{{\rm t}}^{{\rm th}}$ &  $h_{\rm t}^{\rm th}$    \\
\hline
MDI(65)       &  0.070     & 0.213    & 1.0342       &  0.078   & 0.317 & 1.0363    \\
\hline
MDI (72.5)     &   0.064    & 0.213    &  1.0320      &  0.073   &  0.319 & 1.0350   \\
\hline
 MDI (80)      &  0.060     &  0.184   &  1.0310       &  0.069   &   0.295 & 1.0335     \\
\hline
MDI (95)       &   0.050    &  0.074   &   1.0236     &  0.059   &    0.155& 1.0265     \\
\hline
MDI (110)       &  0.044   & 0.031   &  1.0203         & 0.051    &  0.083&   1.0225     \\
\hline
 Sly4            & 0.086   & 0.377   &  1.0184         & 0.098    &  0.578 &  1.0094  \\
\hline
SKI4             & 0.073   &  0.248  &  1.0358         &  0.081   &   0.337 &  1.0378  \\
\hline
Ska              &  0.069  &  0.377  &  1.0409         &  0.079   &   0.530  &  1.0443 \\
\hline
HLPS (soft)       &  0.088  &  0.359  &  1.0394         &  0.098   &   0.455  &  1.0410 \\
\hline
HLPS (stiff)      &  0.079  &  0.415  &  1.0425         &  0.089   &   0.551  &  1.0451  \\
\hline
\end{tabular}
\end{center}
\label{ntptm14}
\end{table}
%%%%%%%%%%%%%%%%%%%%%%%%%%%%%%%%%%%%%%%%%%%%%%%%%%%%%%%%%%%%%%%%%%%%%%%%%%%%%%%
%%%%%%%%%%%%%%%%%%%%%%%%%%%%%%%%%%%%%%%%%%%%%%%%%%%%%%%%%%%%%%%%%%%%%%

\begin{table}[h]
 \begin{center}
\caption{The transition density $n_{\rm t}$ (in fm$^{-3}$),  pressure
$P_{\rm t}$ (in MeV/fm$^3$) and the thermodynamical factor $h_{\rm t}$ obtained from the considered models  by employing
the parabolic approximation. } \label{t:2}
\vspace{0.5cm}
\begin{tabular}{|l|c|c|c|c|c|c|}
\hline
Model &  $n_{\rm t}^{\rm dyn}$  & $P_{{\rm t}}^{{\rm dyn}} $     & $h_{\rm t}^{\rm dyn}$   & $n_{{\rm t}}^{{\rm th}}$     & $P_{{\rm t}}^{{\rm th}}$ &  $h_{\rm t}^{\rm th}$        \\
\hline
MDI(65)       &  0.086     & 0.425    & 1.0389       &  0.097   & 0.594 & 1.0422    \\
\hline
MDI (72.5)     &   0.082    & 0.483    &  1.0397      &  0.094   &  0.728 & 1.0449   \\
\hline
 MDI (80)      &  0.082     &  0.529   &  1.0402       &  0.094   &   0.836 & 1.0469     \\
\hline
MDI (95)       &   0.084    &  0.615   &   1.0396     &  0.099   &    1.079& 1.0497     \\
\hline
MDI (110)       &  0.087   & 0.776   &  1.0426         & 0.105    &  1.406&   1.0556     \\
\hline
 Sly4            & 0.085   & 0.426   &  1.0441         & 0.094    &  0.546   & 1.0462   \\
\hline
SKI4             & 0.082   & 0.356   &   1.0386        & 0.091    &  0.496   & 1.0415   \\
\hline
Ska              & 0.083   & 0.622   &   1.0475        & 0.093    &   0.867  & 1.0524  \\
\hline
HLPS (soft)       &  0.094  &  0.421  &   1.0411        & 0.104    & 0.537  &  1.0430 \\
\hline
HLPS (stiff)      &  0.087  &  0.525   &  1.0453        & 0.097    &  0.694  &  1.0483  \\
\hline
\end{tabular}
\end{center}
\label{ntptm18}
\end{table}
%%%%%%%%%%%%%%%%%%%%%%%%%%%%%%%%%%%%%%%%%%%%%%%%%%%%%%%%%%%%%%%%%%%%%%%%%%%%%%%

%%%%%%%%%%%%%%%%%%%%%%%%%%%%%%%%%%%%%%%%%%%%%%%%%%%%%%%%%%%%%%%%%%%%%%

\begin{table}[h]
 \begin{center}
\caption{The fiducial time scales, the critical frequencies and the critical temperatures  for the MDI model ($L=80$ MeV) for $M=1.4 M_{\odot}$. The corresponding  results of the use of the Tolman VII solution as an approximation have been  included also in a parenthesis for  each case.  } \label{t:4}
\vspace{0.5cm}
\begin{tabular}{|l|c|c|c|c|}
\hline
&  DYN-FE   & THER-FE       &  DYN-PA   & THER-PA      \\
\hline
$\tilde{\tau}_{GR}$       &  -3.72 (-3.67)     & -3.82 (-3.73)    & -3.95 (-3.85)       &  -4.11 (-3.98)      \\
\hline
$\tilde{\tau}_{ee}$     &   40.68 (37.26)    & 33.73 (30.61)    &  26.85 (24.45)      &  22.46 (20.54)      \\
\hline
$\tilde{\tau}_{nn}$      &  94.95 (86.95)     &  77.35 (70.21)   &  60.25 (54.86)       &  49.51 (45.28)        \\
\hline
$f_c^{ee}$       &  823 (834)   & 855 (866)   &   898 (909)         & 936 (946)         \\
\hline
$f_c^{nn}$            & 706 (715)   & 735 (745)   &  775 (785)         &  811 (819)      \\
\hline
$T_c $            & 0.851 (0.916)   & 1.053 (1.133)   &  1.368 (1.464)         &  1.702 (1.802)      \\
\hline
\end{tabular}
\end{center}
\end{table}
%%%%%%%%%%%%%%%%%%%%%%%%%%%%%%%%%%%%%%%%%%%%%%%%%%%%%%%%%%%%%%%%%%%%%%%%%%%%%%%

%%%%%%%%%%%%%%%%%%%%%%%%%%%%%%%%%%%%%%%%%%%%%%%%%%%%%%%%%%%%%%%%%%%%%%
\begin{table}[h]
 \begin{center}
\caption{The fiducial time scales,  the critical frequencies and the critical temperatures for the MDI model ($L=80$ MeV) for $M=1.8 M_{\odot}$. The corresponding  results of the use of the Tolman VII solution as an approximation have been included also in a parenthesis for each case.   } \label{t:4}
\vspace{0.5cm}
\begin{tabular}{|l|c|c|c|c|}
\hline
&  DYN-FE   & THER-FE       & DYN-PA   & THER-PA      \\
\hline
$\tilde{\tau}_{GR}$       &  -0.954 (-0.942)     & -0.967 (-0.950)    & -0.982 (-0.967)       &  -0.997 (-0.975)      \\
\hline
$\tilde{\tau}_{ee}$     &   46.52 (44.33)    & 38.35 (36.30)    &  30.29 (28.94)      &  25.23 (24.00)      \\
\hline
$\tilde{\tau}_{nn}$      &  108.57 (103.47)     &  87.95 (83.25)   &  67.96 (64.92)       &  55.64 (52.90)        \\
\hline
$f_c^{ee}$       &  779 (784)   & 808 (813)   &   846 (851)         & 878 (882)         \\
\hline
$f_c^{nn}$            & 667 (672)   & 695 (699)   &  731 (735)         &  761 (765)      \\
\hline
$T_c $            & 0.191 (0.198)   & 0.235 (0.243)   &  0.302 (0.311)         &  0.368 (0.378)      \\
\hline
\end{tabular}
\end{center}
\end{table}

%%%%%%%%%%%%%%%%%%%%%%%%%%%%%%%%%%%%%%%%%%%%%%%%%%%%%%%%%%%%%%%%%%%%%%%%%%%%%%%

\end{document}